\documentclass[twocolumn]{aastex63}

\submitjournal{ApJ}
\usepackage{amsmath}
\usepackage{ulem}

\newcommand{\Buz}{\textsc{Buzzard}}

\newcommand{\zlower}{$z_{0.0,0.3}$}
\newcommand{\zlow}{$z_{0.3,0.6}$}
\newcommand{\zmid}{$z_{0.6,0.9}$}
\newcommand{\zhigh}{$z_{0.9,1.2}$}
\newcommand{\zhigher}{$z_{1.2,2.5}$}
\newcommand{\nlow}{$\theta\sim1^\circ$}
\newcommand{\nhigh}{$\theta\sim7'$}

\defcitealias{lsstbook}{LSC09}

\shorttitle{Simultaneous estimation of LSS and MW dust}
\shortauthors{Bravo et al.}

\begin{document}

\title{Simultaneous Estimation of Large-Scale Structure\\
and Milky Way Dust Extinction from Galaxy Surveys}

\correspondingauthor{Mat\'ias Bravo}
\email{matias.bravo@icrar.com}

\author[0000-0001-5742-7927]{Mat\'ias Bravo}
\affiliation{Instituto de Astrof\'isica, Pontificia Universidad Cat\'olica de Chile, Avenida Vicu\~na Mackenna 4860, Macul, Santiago, Chile}
\affiliation{International Centre for Radio Astronomy Research, The University of Western Australia, 7 Fairway, Crawley, WA 6009, Australia}

\author[0000-0003-1530-8713]{Eric Gawiser}
\affiliation{Department of Physics and Astronomy, Rutgers University, 136 Frelinghuysen Road, Piscataway, New Jersey, USA}
\affiliation{Center for Computational Astrophysics, Flatiron Institute, 162 5th Ave, New York, NY 10010, USA}

\author[0000-0001-9850-9419]{Nelson D. Padilla}
\affiliation{Instituto de Astrof\'isica, Pontificia Universidad Cat\'olica de Chile, Avenida Vicu\~na Mackenna 4860, Macul, Santiago, Chile}
\affiliation{Centro de Astro-Ingenier\'\i a, Pontificia Universidad Cat\'olica de Chile, Avenida Vicu\~na Mackenna 4860, Macul, Santiago, Chile}

\author[0000-0002-0728-0960]{Joseph DeRose}
\affiliation{Kavli Institute for Particle Astrophysics and Cosmology and Department of Physics, Stanford University, Stanford, CA 94305, USA}
\affiliation{Department of Particle Physics and Astrophysics, SLAC National Accelerator Laboratory, Stanford, CA 94305, USA}

\author[0000-0003-2229-011X]{Risa H. Wechsler}
\affiliation{Kavli Institute for Particle Astrophysics and Cosmology and Department of Physics, Stanford University, Stanford, CA 94305, USA}
\affiliation{Department of Particle Physics and Astrophysics, SLAC National Accelerator Laboratory, Stanford, CA 94305, USA}

\author{The LSST Dark Energy Science Collaboration}



\begin{abstract}
The high cosmological precision offered by the next generation of galaxy surveys hinges on improved corrections for Galactic dust extinction.
We explore the possibility of estimating both the dust extinction and large-scale structure from a single photometric galaxy survey, making use of the predictable manner in which Milky Way dust affects the measured brightness and colors of galaxies in a given sky location in several redshift bins.
To test our method, we use a synthetic catalog from a cosmological simulation designed to model the Vera C. Rubin Observatory Legacy Survey of Space and Time.
At high Galactic latitude ($|b|\gtrsim20^\circ$) and a resolution of $1^\circ$ ($7'$), we predict the uncertainty in the measurement of dust extinction, $E(B-V)$, to be $0.005\ \mathrm{mag}$ ($0.015\ \mathrm{mag}$).
This is similar to the uncertainty of existing dust maps, illustrating the feasibility of our method.
Simultaneous estimation of large-scale structure is predicted to recover the galaxy overdensity $\delta$ with a precision of $\sim$0.01 ($\sim$0.05) at $1^\circ$ ($7'$) resolution.
We also introduce a Bayesian formalism that combines prior information from existing dust maps with the likelihood of Galactic dust extinction determined from the excursion of observed galaxy properties.
\end{abstract}



\section{Introduction}\label{sec:intro}

Distant galaxies are a crucial astrophysical probe, both as an important marker for the mass distribution in the Universe and understanding galaxy formation and evolution.
The advent of the Vera C. Rubin Observatory Legacy Survey of Space and Time \citep[LSST;][hereafter LCS09]{ivezic2019,lsstbook} has the promise to revolutionize several fields in astronomy, as its main survey will combine a sky coverage similar to the widest existing ground-based surveys\footnote{$\sim20,000\ \deg^2$, compared to $\sim15,000\ \deg^2$ from the Sloan Digital Sky Survey \citep{alam2015}.} with an improved depth compared to the deepest current wide-area surveys\footnote{$r\lesssim28$ after ten years, compared to $r\lesssim26$ from the Hyper Suprime-Cam Subaru Strategic Survey \citep{aihara2018}.}.
As the LSST will discover billions of galaxies, aspects of the analysis such as Large-Scale Structure (LSS) that have previously been limited by statistics will become systematics-limited.

This has motivated significant efforts to predict the effect of these systematics and how to minimize them.
For example, \citet{awan2016} studied the impact on the effective survey depth of different dithering strategies for the LSST, finding that it is possible to reduce these errors to a level where Milky Way (MW) dust extinction becomes the dominant large-scale structure systematic at larger angular scales.
This means that accurate mapping of this foreground screen becomes a crucial requirement to maximize the use of LSS measurements for cosmology studies with the LSST.

In what has become a seminal work, Schlegel, Finkbeiner, and Davis \citep[][hereafter SFD98]{schlegel1998}\defcitealias{schlegel1998}{SFD98} released a full-sky dust map of the MW, made from direct measurement of the infrared emission of galactic dust to derive its column density, and from that a dust extinction map assuming a fixed dust law.
They used the infrared maps of the sky from two space telescopes, the Diffuse Infrared Background Experiment \citep[DIRBE;][]{hauser1991} and the IRAS Infrared Survey Atlas \citep[ISSA;][]{issabook}, from the Cosmic Background Explorer \citep[COBE;][]{boggess1992} and Infrared Astronomical Satellite \citep[IRAS;][]{irasbook}, respectively.
Due to the characteristics of those observations, the resulting map is zero-point calibrated at a resolution of $\sim1^\circ$, while the higher frequencies (from $\sim1^\circ$ to $\sim4'$) have only a relative calibration.

Due to the immense relevance of accounting for MW dust extinction for extragalactic astronomy, the subsequent years witnessed a large effort to test the \citetalias{schlegel1998} dust map using different observables and catalogs.
This includes the US Naval Office - Precision Measuring Machine \citep[USNO-PMM;][]{monet1996} stellar counts \citep{cambresy2001}; Sloan Digital Sky Survey \citep[SDSS;][]{york2000} galaxy counts \citep{yasuda2007}; SDSS passively-evolving galaxy colors \citep{peek2010}; SDSS stellar Main-Sequence turn-off \citep{schlafly2010}; SDSS M-dwarf stellar spectra \citep{jones2011}; SDSS stellar spectra \citep{schlafly2011}; SDSS quasar, central galaxies and LIRG colors \citep{mortsell2013}; and SDSS + South Galactic Cap $u$-band Sky Survey \citep[SCUSS;][]{zhou2016} galaxy counts and colors \citep{li2017}.
New maps have also been released, using Wide-field Infrared Survey Explorer \citep[WISE;][]{wright2010} Polycyclic Aromatic Hydrocarbon emission measurements \citep{meisner2014}, IRAS+Planck \citep{planck2014i} direct IR imaging \citep{planck2014xi,meisner2015,planck2016xxix,planck2016xlviii}, Panoramic Survey Telescope and Rapid Response System \citep[Pan-STARRS1;][]{chambers2016} stellar photometry \citep{schlafly2014}, Pan-STARRS1+Two Micron All Sky Survey \citep[2MASS;][]{skrutskie2006} stellar spectra \citep{green2015,green2018}, HI $4\pi$ \citep{hi4pi2016}, Galactic HI emission \citep{lenz2017}, and Pan-STARRS1+2MASS+Gaia \citep{gaia2016} stellar spectra \citep{green2019}.
Though the conclusions from the comparisons with the \citetalias{schlegel1998} map vary, there seems to be a broad agreement that it modestly under- and over-estimates the extinction at high and low galactic latitudes, by $\sim20\%$ and $\sim50\%$ respectively \citep[e.g., ][]{yasuda2007,peek2010,mortsell2013,li2017}.
A factor that adds complexity to the use and evaluation of these maps is that most of them are contaminated by (unaccounted) background LSS, seen as over- or under-estimation of dust extinction correlated with overdensity \citep{chiang2019}.

Currently, there are a handful of methods to handle systematics like MW dust for LSS science \citep[see][for an overview of existing and proposed methods]{elsner2016,weaverdyck2020}, which all depend on including a map such as that of \citetalias{schlegel1998} to correct either overdensity measurements or their power spectra.
A galaxy overdensity field, pixelized on various angular scales, can be used directly for cosmological measurements of clustering and can also serve as weights for clustering measurements of redshift-space distortions and baryon acoustic oscillations.

Weights are commonly used in the literature \citep{hawkins2003,alam2017} and are designed to remove systematic biases arising from environment-dependent completeness effects, such as fiber collisions, and to reduce noise in the measured clustering.
The survey depth is usually calculated by taking into account the effect of Galactic dust on survey magnitude limits.
This calculated depth enters clustering calculations through the use of random catalogs, produced with angular variations applied to the depth map such that they feature the same angular dependence \citep{hawkins2003}.

However, galaxy weights rarely include the effects of dust on projected overdensities, except in some cases at the level of defining the target sample in redshift surveys (e.g. \citealt{ross2017}).
This correction could be particularly helpful for photometric surveys; for instance, \citet{salazar2017} use clustering tomography to constrain cosmological parameters without correcting for the angular dependence of Galactic dust.
Given the concern about LSS contamination of existing dust maps noted earlier, the application of any such methods is bound to propagate the systematics present in the dust map of choice.

To this end, here we present a new method to simultaneously derive an MW dust map and correct the observed LSS for its effect, using data from photometric galaxy surveys.
Our approach takes advantage of the fact that extinction and reddening (due to MW dust) are related phenomena, connected by a choice of dust law \citep[e.g.,][]{fitzpatrick1999,calzetti2000,yuan2013,maiz2014}.
This allows the use of statistics of the observed magnitudes and colors of extragalactic objects to identify deviations from the expected intrinsic distribution of these properties, which we show can be used to put constraints on the MW dust extinction.
\citet{yasuda2007} and \citet{li2017} point out that the construction of a dust map from SDSS galaxy catalogs with strong statistical power is impossible at resolutions similar to the existing ones ($\lesssim1^\circ$), and while the Dark Energy Survey \citep[DES;][]{des2005} is capable of detecting galaxies $\sim10$ times dimmer than SDSS, its sky coverage ($\approx5000$ deg$^2$) is significantly smaller.
The LSST is deeper than DES and covers as much area as SDSS, making it a particularly promising source for a new dust map.
In this work, we study the viability of such an undertaking using a synthetic galaxy catalog generated from a cosmological simulation.

In \S\ref{sec:synth_cats}, we describe the numerical simulation and the phenomenological model used to obtain simulated LSST lightcone (LC).
We describe our method to apply MW dust extinction on this mock catalog and then to recover an MW dust map, along with the results from our application of the method to the LC in \S\ref{sec:EBV_maps}.
Using this recovered map, we present in \S\ref{sec:delta_maps} the method to recover the input Large Scale-Structure and the results from the simulations.
In \S\ref{sec:discussion}, we present a discussion and introduce a Bayesian expansion of our method to refine an existing dust map using deep galaxy surveys while simultaneously measuring LSS.
The summary of our results is laid out in Section \S\ref{sec:summary}.


\section{Synthetic galaxy catalog}\label{sec:synth_cats}

To assess the possibility of constructing dust maps with future photometric galaxy surveys, specifically LSST for this work, we need realistic synthetic catalogs containing information about the relevant galaxy properties and their evolution.
For this, we use available cosmological numerical simulations which follow the evolution of the dark matter (DM) distribution, populated with galaxies using a phenomenological model for the relation between galaxies and dark matter.

N-body DM-only simulations follow gravitational interactions to obtain a final matter distribution and, therefore, halo population.
This is less expensive computationally than a full hydro-dynamical model, allowing for the simulation of large cosmological volumes ($\sim0.1$-$10$ Gpc$^3$), which is required for constructing synthetic LC for surveys with large footprints and deep photometry.
The synthetic galaxy catalog we use in this work was built from the \textsc{Chinchilla}-400 N-body DM-only simulation \citep{lehmann2017,derose2019a}, with a box size of $400\ \mathrm{h}^{-1}\mathrm{Mpc}$, $2048^3$ particles, and a $5.5$ $\mathrm{h}^{-1}\mathrm{kpc}$ force softening kernel, and has been referred to elsewhere as the \Buz\ highres simulation \citep{malz2018}.
This simulation provides the spatial and clustering information of halos, which can be used as input to generate a synthetic galaxy catalog.
This choice complies with the requirements for the structure described by galaxies to be fully non-linear, and the volumes spanned by the LCs to be large enough to cover a sizeable fraction of the LSST.
The first requirement, in particular, makes us prefer N-body simulations over approximate methods such as those used to produce hundreds of mock catalogs for covariance matrix estimation such as QPM Mocks \citep{white2014}.

There are multiple methods to populate DM-only simulations with galaxies, such as halo occupation distributions \citep[HOD; e.g.,][]{berlind2002}, sub-halo abundance matching \citep[SHAM; e.g.,][]{vale2004,kravtsov2004,conroy2006}, and semi-analytic models \citep[SAM; e.g., ][]{kauffmann1993,cole1994,somerville1999,cole2000}; \cite{wechsler2018} provides an up-to-date review of these different approaches.
The galaxy catalog we use in this work was generated using the same SHAM method used to produce the \Buz\ flock LCs for DES \citep{derose2019b}, and also described in Appendix A of \citet{wechsler2021}.
It does not produce a full quarter sky like the \Buz\ simulation described in \cite{derose2019a,derose2021}, but extends to a higher redshift than these simulations.
For this LC, sub-halos were identified down to 10 particles, to obtain galaxies down to $r\sim29$\footnote{This LC is only magnitude-complete at $z\gtrsim0.4$, due to the mass resolution of the parent simulation.}.
Appendix \ref{app:other_sims} offers a comparison between this LC and a comparable (but not for LSST) LC made with the \textsc{GALFORM} semi-analytic model \citep{bower2006}.

Galaxy magnitudes are added to the simulations using a variant of the method described in Section 5 of \citet{wechsler2021}.
Galaxy SEDs from an SDSS training set are assigned to each simulated galaxy using a relationship between galaxy SED, absolute magnitude, and projected distance to the galaxy's fifth nearest neighbor.
Each synthetic galaxy is assigned the SED from a galaxy in the training set in the same absolute magnitude and distance bin, with the probability of drawing a red galaxy defined as a function of redshift and luminosity (see appendix E.2 of \citet{derose2019b}).
SEDs are represented using five \textsc{kcorrect} templates \citep{blanton2007}, which are constructed by fitting to $griz$ photometry for galaxies selected from SDSS Main Galaxy Survey \citep{strauss2002}.
Simulated galaxies are imbued with the \textsc{kcorrect} template coefficients of the observed galaxy assigned to them, allowing for the generation of photometry in all desired bands.
The SEDs were shifted accordingly to produce the correct observer-frame magnitudes for the light cone.

It is important to note that the use of SDSS catalogs throughout the construction of the \Buz\ highres LC leads to two caveats to its use, as SDSS galaxies are not fully representative of the dimmer and younger galaxies that the LSST will probe, and that the variance in the rest-frame $u$ band (not part of the training set) is mildly underestimated \citep{wechsler2021}.
Since the size of this simulation is not large enough to reach the LSST redshift detection limit, the box needed to be replicated, taking advantage of its periodic boundary conditions.
As described by \citet{merson2013a}, this introduces structure repetition if the boxes are not reoriented, or mass field discontinuities if rotations are applied.
The \Buz\ highres LC adopts the former approach, choosing a sky footprint for the LC of $398.5$ deg$^2$ as a balance between structure repetition and LC volume.

Finally, the survey magnitude and redshift limits ($r\leq29$ and $z\leq8.7$) were applied to the LC, to go slightly beyond the designed LSST coadded depth of $r\leq27.7$ (table 1.1 of \citetalias{lsstbook}).
We should note, however, that the \Buz\ highres LC suffers magnitude incompleteness due to the mass resolution of the underlying DM-only simulation, with the completeness limit being $r\approx25$-$27$, depending on redshift.
While the more recent \textsc{CosmoDC2}-based LSST DESC Data Challenge 2 Simulated Sky Survey \citep[DC2SSS;][]{korytov2019,lsstdesc2020} does not suffer from such magnitude incompleteness, its current implementation of random positions for sub-resolution galaxies would negatively impact testing our method, which depends on whether the simulation broadly reproduces the environment-brightness-color relations exhibited by real galaxies.

To mimic observed magnitudes, we add errors to the photometry and redshifts of the synthetic LC, following the expected errors in the LSST main survey (see \S1.6.1 and \S3.8.1 of \citetalias{lsstbook}).
For the photometry, we add noise via
\begin{equation}
    m_{i,\mathrm{obs}}=-2.5\log\left(10^{-0.4m_{i,\mathrm{sim}}}+\varepsilon_{f,i}\right),\label{eq:1}
\end{equation}
with $\varepsilon_{f,i}$ drawn from a normal distribution with $\mu=0$ and $\sigma=0.2\times10^{-0.4m_{f,\lim}}$, where $m_{f,\lim}$ is the $5\sigma$ coadded depth for each filter $f$.
We assign a magnitude of $99$ to all cases where $10^{-0.4m_i}+\varepsilon_{f,i}\leq0$.
For the redshifts, we divide the galaxies between the ``gold" sample ($i<25.3$, see \S3.7.2 of \citetalias{lsstbook}) and those outside that selection.
For those in the ``gold" sample, we modify the true redshifts as
\begin{equation}
    z_i=z_i+(1+z_i)\varepsilon_{z,i},\label{eq:2}
\end{equation}
with $\varepsilon_{z,i}$ drawn from a normal distribution with $\mu=0$ and $\sigma=0.02$.
For the rest of the galaxies, we use a similar prescription, with $\sigma$ a linear function of the $i$-band magnitude, following
\begin{equation}
    z_i=z_i+(1+z_i)\varepsilon_{z,i}^*,\label{eq:3}
\end{equation}
with $\varepsilon_{z,i}^*$ drawn from a normal distribution with $\mu=0$ and $\sigma=0.02+0.01\frac{i-25.3}{\mathrm{mag}}$.
To model the redshift-calibrated sample of the LSST (see \S3.8.1 of \citetalias{lsstbook}), we apply a magnitude selection of $r<26.0$, which also removes the incompleteness issue at higher redshifts ($z\gtrsim0.5$), though this remains a limitation of the synthetic LC at lower redshifts.


\section{Measuring dust extinction from galaxy properties}\label{sec:EBV_maps}

We now describe a method to measure the MW dust extinction through the use of statistics of the observed magnitudes and colors of galaxies.
The basic procedure is the following:

\begin{enumerate}
    \item Divide the range of redshift covered by the survey into redshift bins.
    \item Determine a relation between dust extinction and the changes induced by it on the median magnitude and color of galaxies in each redshift bin. This can be done with simulations or with the observed distribution of galaxy properties in each bin. \S\ref{sec:dust_vector} describes the method in this work.\label{list:method_steps_11}
    \item Divide the area of the sky covered by the survey into angular pixels. Combined with the redshift binning, this creates volumetric pixels (voxels)\footnote{Following the naming convention used in \citet{green2019}.}.
    \item Calculate the median observer-frame apparent magnitudes and colors of the galaxies for each voxel.\label{list:method_steps_12}
    \item Subtract the median value of each property across the voxels in the same redshift bin, defining $\Delta_\mathrm{magnitude}$ and $\Delta_\mathrm{color}$.
    \item For each voxel in this $\Delta_\mathrm{magnitude}$-$\Delta_\mathrm{color}$ space, calculate the closest point to the curve from the relation defined in step \ref{list:method_steps_11} to find the corresponding dust extinction value for the voxel.\label{list:method_steps_13}
    \item Calibrate the $y$-intercept of the recovered relative maps, by either matching to a chosen statistic from an existing map (within the survey area), or using an independent $E(B-V)$ measurement to calibrate against (like the Mg$_2$\footnote{Magnesium index described by \citet{faber1977}}-color relation used by \citetalias{schlegel1998}).
    \item Combine the voxels sharing the same sky pixel using e.g., an inverse-variance weighted average, to produce an $E(B-V)$ map.
\end{enumerate}

The main reason to divide the survey into redshift bins is the strong evolution of observed colors with redshift, which would induce a high dispersion in color values of the voxels, impacting our capability to separate the effect from dust.
This has the added benefit of enabling multiple independent measurements, which combined will lead to decreased errors.
In this work we use five bins: $0<z<0.3$ (from this point referred as \zlower), $0.3<z<0.6$ (\zlow), $0.6<z<0.9$ (\zmid), $0.9<z<1.2$ (\zhigh) and $1.2<z<2.5$ (\zhigher).
This choice was made trying to balance the number of galaxies in each redshift bin while keeping the $1.4<z<2.5$ range in a single bin, as it is expected that in this range LSST will suffer from larger photometric redshift errors (see \S3.8 of \citetalias{lsstbook}).

We note that step \ref{list:method_steps_12} above implies that for any given resolution, we are only using the voxels with at least one detected galaxy.
This puts a practical limit on the highest resolution achievable, as some of the voxels inside the survey volume will be devoid of galaxies below a certain pixel and bin size, and the statistics notably worsen as the median number of galaxies per voxel approaches 1.
We choose to test two resolutions: \nlow\ and \nhigh, where we use $\theta$ to denote angular resolution.
We choose these resolutions as they roughly match the two resolutions of the \citetalias{schlegel1998} map, with $7'$ also being similar to the maps by \citet{schlafly2014,planck2016xxix,lenz2017}.
For this, we use the Hierarchical Equal Area isoLatitude Pixelization package \citep[HEALPix;][]{healpix}, where these resolutions correspond to values of $N_\mathrm{nside}$ of $64$ and $512$.
The larger the number of galaxies, the smaller the dispersion in their properties, allowing a better estimate of the foreground dust but with a decreased spatial resolution for the resulting map.

\begin{figure}
    \epsscale{1.1}
    \plotone{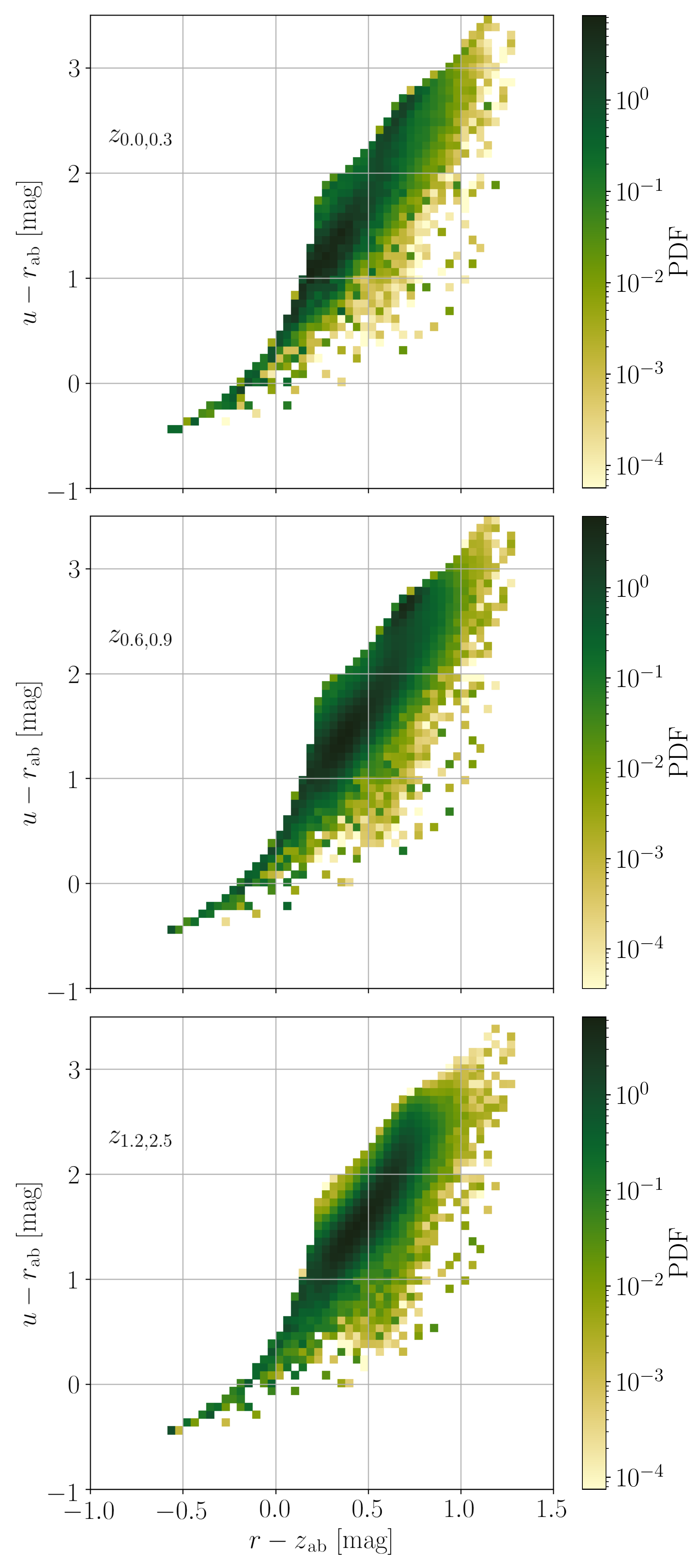}
    \caption{Color-color diagrams from the LSST \Buz\ highres lightcone. Colors are in rest-frame absolute magnitudes. Each panel shows a different redshift bin, from top to bottom: \zlower, \zmid\ and \zhigher.}
    \label{fig:col_col_z_ap}
\end{figure}

While we choose to use a single color in this work, it is relevant to remark upon the importance of using multiple colors in any real-world implementation of this method.
Dust law variations, both between different laws \citep[e.g.,][]{fitzpatrick1999,calzetti2000,yuan2013,maiz2014} and the choice of parameters in them, have a non-negligible impact on the conversion from $E(B-V)$ to reddening in any pair of bands, and any single color is not sensitive to such variations.
This is not a problem in the work presented here, as we consistently use the empirical results from \citet{yuan2013}\footnote{Third column of their table 2.
While SDSS and LSST filters do not have identical spectral responses, for the purpose of this work this is a good enough approximation.} both to simulate the dust screen and to recover it, but it is a significant issue to be addressed for the implementation of our method.

Figure \ref{fig:col_col_z_ap} shows the observer-frame color-color distribution from our LC for three redshift bins.
While the distribution strongly evolves with redshift, for every redshift bin, the distribution is highly correlated.
Since one color dimension encodes most of the information on the synthetic LCs, we expect the use of multiple colors to provide only second-order improvements, with the primary advantage of using multiple colors being the added ability to account for dust law variations.

While for our method we do not explicitly need to use an existing dust map, as LSST will produce only photometric redshifts, the fact that we are binning the galaxies by redshift implies the requirement of a dust map to use for the redshift measurements.
One solution would be to start with the assumption of no dust extinction for a first photometric redshift measurement, use it to generate the extinction map, and then iterate between these two steps until convergence.
However, the most sensible choice is to use one of the available dust maps to correct the photometry before a first photometric redshift measurement, iterating these measurements only if necessary.

\begin{figure}
    \epsscale{1.1}
    \plotone{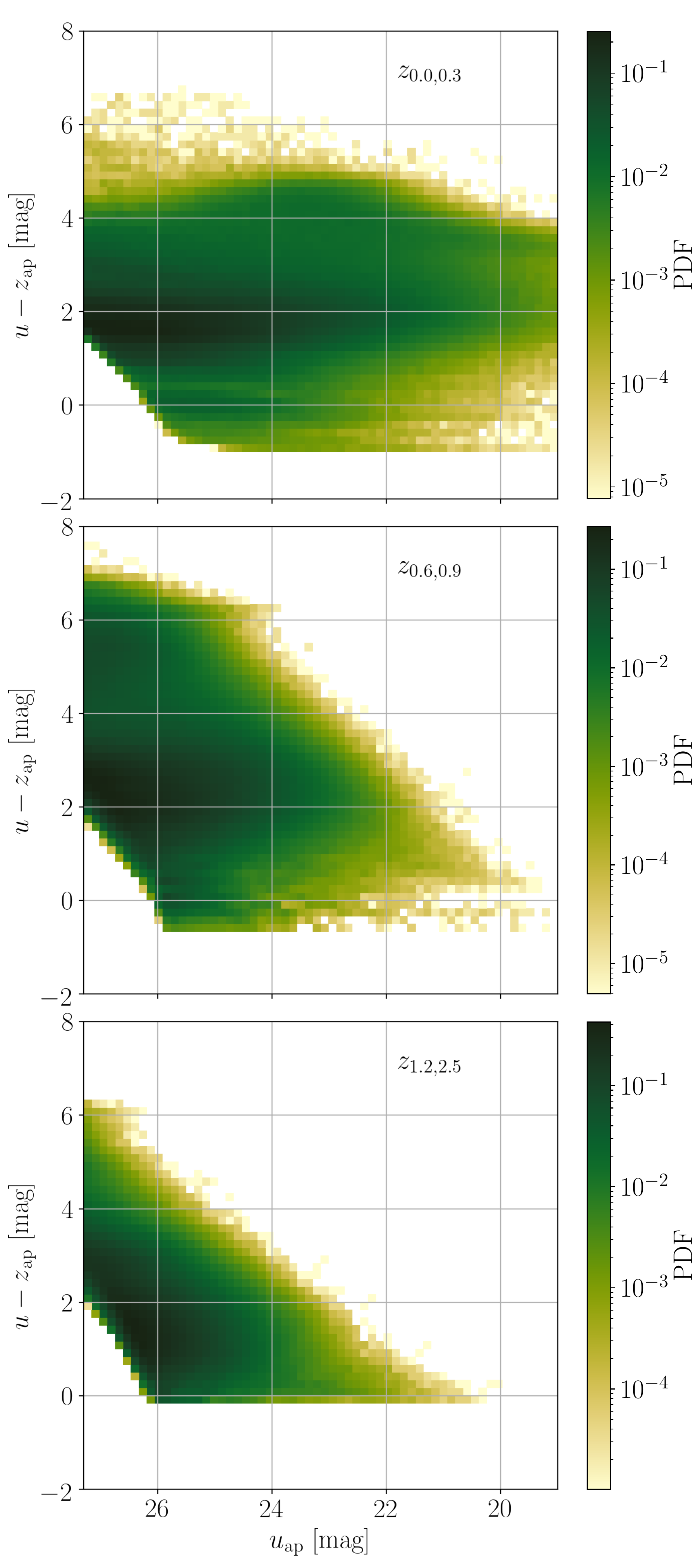}
    \caption{Color-magnitude diagrams of the galaxies from the synthetic LC used in this work. Each panel shows a different redshift bin, from top to bottom: \zlower, \zmid, and \zhigher. We display $u$ and $u-z$ as these are the filters that we use for our method (see \S\ref{sec:filters}).}
    \label{fig:mag_col_z_ap}
\end{figure}


\subsection{Choice of filters}\label{sec:filters}

Since dust extinction depends on the wavelength of the light that it absorbs, the selection of the filters is highly relevant.
For the magnitude component (of the color-magnitude space), it is straightforward that choosing the bluest filter available would maximize the dust signal.
For the color component, to amplify the signal, the best choice is pairing the bluest filter (same as with magnitudes) with the reddest filter unaffected by dust re-emission.
In this work, we will exclude the $y$ filter from consideration, due to its low throughput (see \S2.4.1 of \citetalias{lsstbook}), which leads to the $z$ filter as the choice for the red part of the color index.
While the $u$ filter is also expected to have a low throughput, it would be the filter most affected by extinction, so the choice for the magnitude and blue part of the color becomes a choice between $u$ or the more sensitive $g$.
We tested both $u$-$z$ and $g$-$z$ pairs, and found no qualitative difference, with the latter leading to an increase in noise in the recovery of both MW dust and LSS (by a factor $<2$).
In this work, we choose to present our results for $u$-$z$, as the better performance should more closely follow the results when using all bands.

LSST object detection will be performed on stacked images across all bands, which given the depth in each filter means that the detection will be dominated by the $gri$ filters, which we model by applying an $r<26$ selection to the synthetic LC.
This means that some detected galaxies will have low SNR photometry in the $u$ (or $z$) filter.
For this reason, in addition to the selection on $r$, we add a $2\sigma$ threshold for both $u$ and $z$ magnitudes to improve the dust measurement, which translates to limiting magnitudes of $u=27.3$ and $z=27.2$.
Figure \ref{fig:mag_col_z_ap} shows the observed color-magnitude distribution resulting from applying this multi-filter selection into the synthetic LC at three different redshift ranges.

The LSST Science Requirements Document\footnote{\url{https://www.lsst.org/scientists/publications/science-requirements-document}} promises photometric calibration at the root mean square level of $0.02$ $\mathrm{mag}$ in $u$ and $0.01$ $\mathrm{mag}$ in all other bands.
If this is achieved, it implies that a false dust signal due to photometric calibration errors should not generally be larger than $A_u=0.04$ $\mathrm{mag}$ or $A_z=0.02$ $\mathrm{mag}$, which corresponds roughly to $E(B-V)\sim0.008$ $\mathrm{mag}$ from $u$ and $E(B-V)\sim0.016$ $\mathrm{mag}$ from $z$, using the model-free $E(B-V)$-to-$A_{\{u,r,z\}}$ conversions from \citet{yuan2013}.
To mimic a dust signal, the two bands must fluctuate together (in magnitude and color), so in general, the induced errors in dust recovery will be less than implied by this back-of-envelope calculation.

\begin{figure}
    \epsscale{1.1}
    \plotone{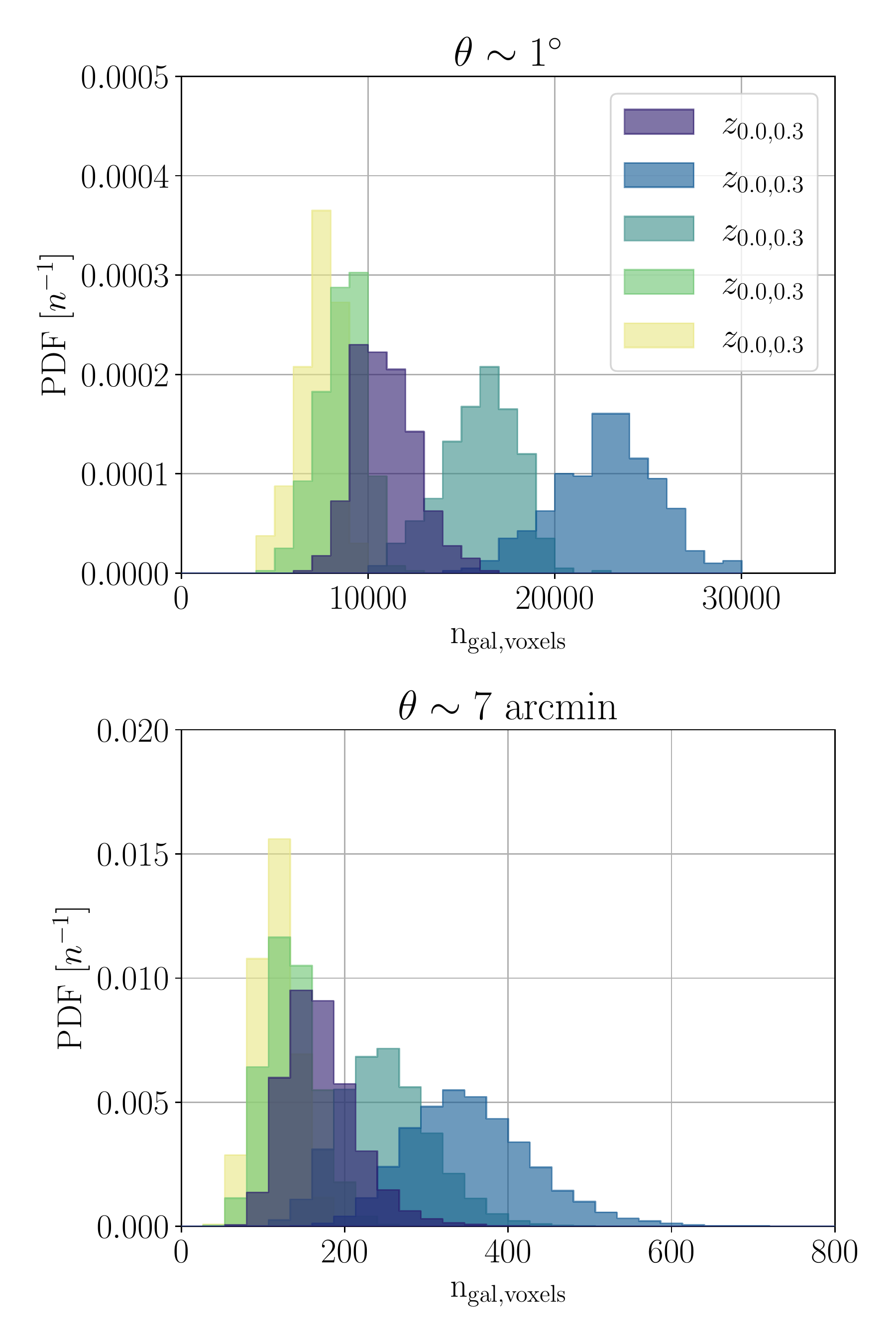}
    \caption{Distribution of voxel number counts, colored by redshift. Upper panel for our low resolution (\nlow) and lower for the high resolution (\nhigh).}
    \label{fig:count_dist}
\end{figure}

Figure \ref{fig:count_dist} shows the distribution of number counts for the voxels assembled at the two chosen resolutions for the synthetic LC.
At \nlow\ median number counts for each redshift bin vary by no more than a factor of 3, with even the voxels with the lowest count in \zhigher\ providing a signal-to-noise ratio (SNR) $\sim70$, with most of the voxels with SNR$>100$, which ensures strong statistics.
The move to \nhigh\ has a significant impact on number counts, with the median galaxy count for \zlower, \zhigh, and \zhigher\ being below 200 galaxies per voxel (SNR$\approx15$), which should still be good enough to ensure a good dust map recovery, albeit with larger errors.


\subsection{Converting magnitude-color to $E(B-V)$}\label{sec:dust_vector}

Step \ref{list:method_steps_11} of the method we describe in \S\ref{sec:EBV_maps} requires establishing the relation between the MW dust extinction and the magnitude and color we measure for each voxel.
Because each voxel has a distribution of colors and magnitudes, and once dust extinction is added, the distributions are further modified by detection limits, the relation between magnitude/color medians and $E(B-V)$ need to be estimated using an empirical approach.
To measure this relation we take each of the whole redshift bins (not subdivided into voxels) and apply a dust screen, with values applied in the range 0 to 0.2 $E(B-V)$.
We then use the transformations from \citet{yuan2013} to convert $E(B-V)$ into $A_u$, $A_r$, and $A_z$ values, by which we attenuate each galaxy.
To this new photometry, attenuated by simulated MW dust, we apply the same magnitude selections detailed in \S\ref{sec:filters} (i.e., no MW dust correction is used).
Finally, we calculate the change in median $u$ and $u-z$ for each redshift bin, relative to the medians of said values with no dust screen ($E(B-V)=0$).

\begin{figure}
    \epsscale{1.1}
    \plotone{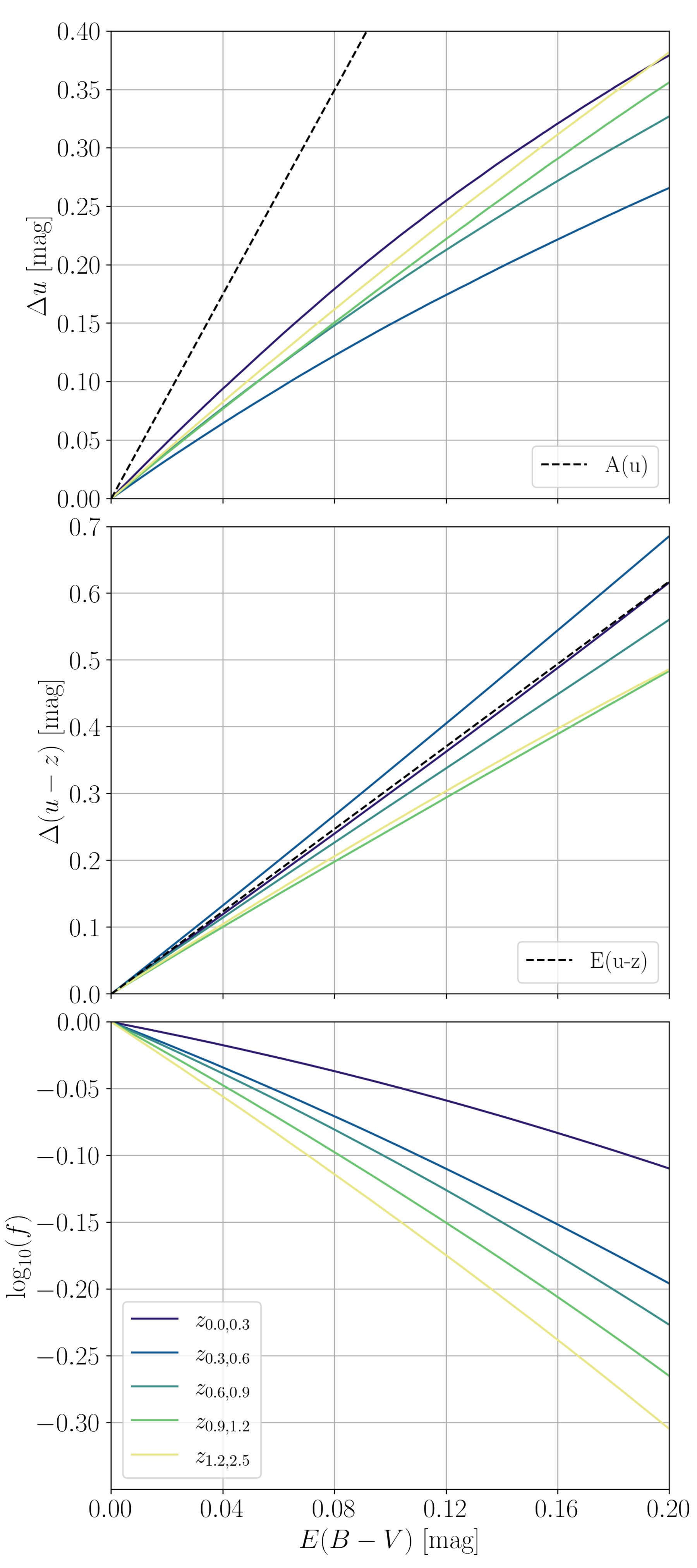}
    \caption{Change of the measured galaxy properties as a function of the added dust screen. Each row shows the results for the considered properties ($u$, $u-z$, and $\log(1+\delta)$). Color lines indicate the measured change for the different redshift bins. The black dashed lines show  
    the simplified assumption that the changes in median magnitude or color are determined precisely by the attenuation or reddening of ($u-z$) of the dust law for a given amount of E(B-V). Note that while we present the effect of dust on galaxy overdensities, only the results for $u$ and $u-z$ are used for dust recovery, with the third panel coming into play for LSS reconstruction.}
    \label{fig:dust_to_quantities}
\end{figure}

While using this exact method for observed data would still require the use of simulations, a near equivalent for applying this method to observations could be achieved by using a slightly deeper sample from a known low-extinction region of the sky.
Another alternative would be to use a broader region (even the full survey), which will be more affected by dust, measure a first dust map, correct the survey by it and then repeat this process iteratively.
Either method would avoid any biases from a possible mismatch in galaxy properties between observations and simulations.

\begin{figure}
    \epsscale{1.1}
    \plotone{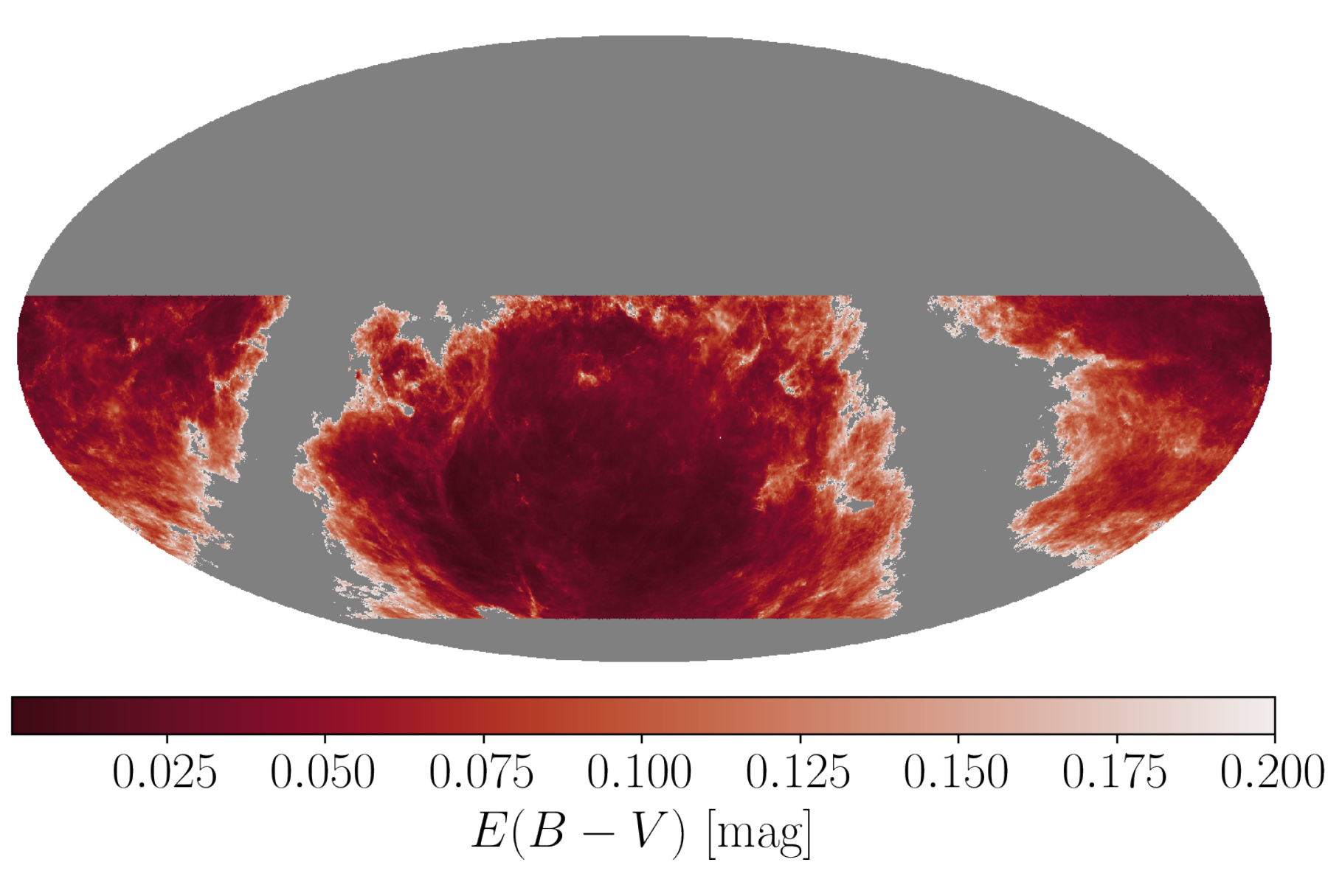}
    \caption{A selected region of the \citet{schlegel1998} dust extinction map, shown in a Mollweide projection, with declination ranging from $90^\circ$ to $-90^\circ$ from top to bottom, and right ascension from 0 to 24 h from left to right. $E(B-V)$ is shown for the areas that we have selected to draw random samples from, $E(B-V)<0.2$ and $-70^\circ<$dec$<12.5^\circ$, with the pixels outside this range shown in gray.}
    \label{sfdcutmap}
\end{figure}

The top and middle panels of Figure \ref{fig:dust_to_quantities} show the changes in the median magnitude and color of galaxies in the synthetic LC as a function of the applied MW dust screen, for different redshift bins.
As a reference point, we also show the change in quantities expected if one makes the simple assumption that the median galaxy magnitude and color will change exactly by the $A_u$ and $E(u-z)$ values from the added dust screen.
While for galaxy colors that simple assumption is broadly consistent with our results, there are still differences of up to a factor of $\sim1.5$ between redshift bins.
This could lead to errors of up to $\sim20\%$ in $E(B-V)$ measured from $u-z$ in \zhigh\ and \zhigher.
The discrepancy between that simple assumption and our results is much stronger in magnitude.
This serves to highlight the danger of such a simple assumption and should not be an unexpected result, as the only way for that assumption to be true is if no galaxies go below the selection threshold due to the dust screen, which is not a realistic scenario.

\begin{figure*}
    \epsscale{1.1}
    \plotone{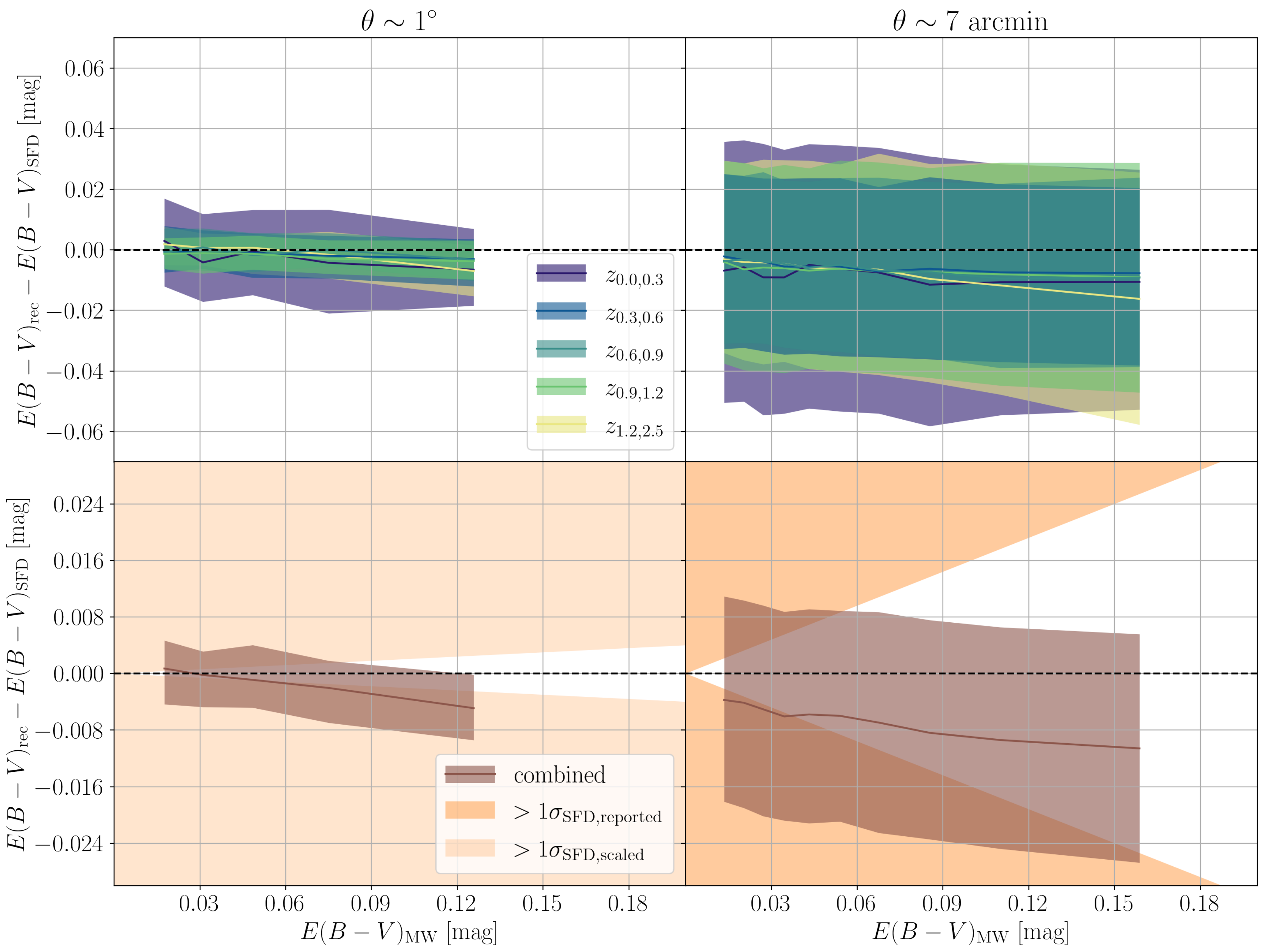}
    \caption{Residuals between the recovered and input dust maps, $E(B-V)_\mathrm{rec}$ and $E(B-V)_\mathrm{SFD}$ respectively. The solid lines show the running median in bins of an equal number of galaxies, the shaded areas show the 16-84\% range. Columns show the results for each resolution. Top row: $E(B-V)$ values from voxels, colored by redshift bin. Bottom row: values for sky pixels, from the inverse-variance weighted combination of the voxels sharing the same pixel, with the orange-shaded area showing the errors from the \citetalias{schlegel1998} map (at the map native resolution on the right, naively scaled to \nlow on the left).}
    \label{fig:EBV_recovery}
\end{figure*}


\subsection{Recovery of simulated Milky Way dust screen}\label{sec:EBV_recover}

\begin{figure*}
    \epsscale{1.1}
    \plotone{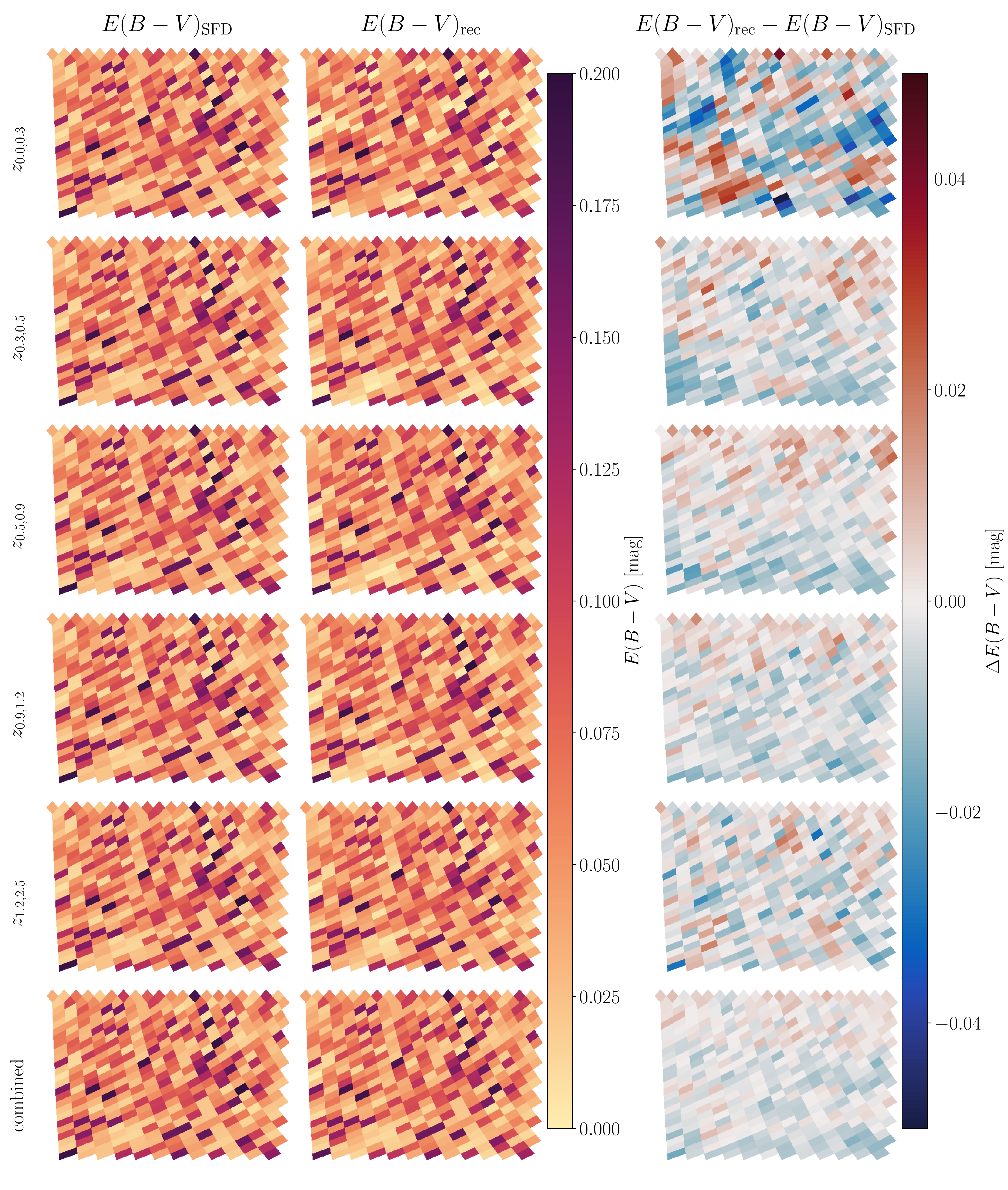}
    \caption{Cartesian view of the input, recovered, and residual maps for MW dust extinction; $E(B-V)_\mathrm{SFD}$, $E(B-V)_\mathrm{rec}$ and $E(B-V)_\mathrm{rec}-E(B-V)_\mathrm{SFD}$ respectively. The top five rows show different redshift bins, while the bottom row shows the combined results. All panels on the left column show the same data, they are repeated for easier comparison with the middle column panels. Note that all panels in the left column display the same data, as a single MW dust screen is applied to the LC. This is done for easier comparison with the recovered MW dust maps in the middle column panels.}
\label{fig:EBV_map_n6}
\end{figure*}

With a model of how dust affects the observed galaxy properties, we test our method by adding a dust screen to the synthetic LC, randomly sampled from the \citetalias{schlegel1998} map.
Following \citet{lochner2018}, we restrict the values from which we sample to the intersection of $E(B-V)<0.2$ mag and $-70^\circ<\mathrm{dec}<12.5^\circ$, shown in Figure \ref{sfdcutmap}.
For the $y$-intercept calibration, we choose to match the medians of the recovered maps at each redshift bin to that of our sample region from the \citetalias{schlegel1998} map.

Figure \ref{fig:EBV_recovery} compares our recovered dust map, $E(B-V)_\mathrm{rec}$ to the input map, $E(B-V)_\mathrm{SFD}$, at the two resolutions of \nlow\ and \nhigh, with Figure \ref{fig:EBV_map_n6} showing this data in a Cartesian map visualization; Table \ref{tab:EBV} shows a summary of the statistics from Figure \ref{fig:EBV_recovery}.
For both resolutions, the results from each redshift bin are in good agreement with the other bins, with only \zlower\ displaying an increased scatter in $E(B-V)_\mathrm{rec}-E(B-V)_\mathrm{SFD}$ compared to the rest.
There is a small but visible negative slope in all redshift bins and resolutions, which we believe comes from how we have implemented step \ref{list:method_steps_13}.
To simplify our calculation we are measuring distances with a fixed metric, while the correct method would use differential geometry, as Figure \ref{fig:dust_to_quantities} shows that the relations of either $\Delta u$ and $\Delta u-z$ are not linear with $E(B-V)$.
There is also a small $y$-intercept offset remaining, which comes from the different distribution of values between recovered and input maps.

Even with this shortcoming, it is remarkable that at \nlow\ by combining all redshift bins we can recover the input dust screen to $|E(B-V)_\mathrm{rec}-E(B-V)_\mathrm{SFD}|\lesssim0.008$ mag.
Assuming that we can further reduce the $y$-intercept and slope differences, the results displayed in Table \ref{tab:EBV} show that the scatter on the recovery could be as small as $|E(B-V)_\mathrm{rec}-E(B-V)_\mathrm{SFD}|\lesssim0.005$.
As expected, the increase in Poisson noise from the increase in resolution going to \nhigh\ translates into a larger scatter (by a factor of $\sim3$).
The slope in the residuals appears to be resolution-independent, as it is consistent with that from \nlow\, which is evidence that the source is how we measure distances in the color-magnitude space, though this is not the case for the $y$-intercept offset, as it increases by the move to \nhigh.

\begin{deluxetable}{lcc}
    \tablecaption{Standard deviation of $E(B-V)_\mathrm{rec}-E(B-V)_\mathrm{SFD}$ (in magnitudes) of the voxels, at each redshift and the final combined value.
    \label{tab:EBV}}
    \tablehead{\colhead{}&\colhead{\nlow}&\colhead{\nhigh}}
    \startdata
    \zlower           & 0.01482 & 0.04429\\
    \zlow             & 0.00765 & 0.02956\\
    \zmid             & 0.00713 & 0.02928\\
    \zhigh            & 0.00609 & 0.03524\\
    \zhigher          & 0.00767 & 0.03599\\
    \textbf{combined} & \textbf{0.00465} & \textbf{0.01547}\\
    \enddata
\end{deluxetable}

Another interesting feature is that the scatter in $E(B-V)_\mathrm{rec}-E(B-V)_\mathrm{SFD}$ from our method does not appear to depend on $E(B-V)_\mathrm{SFD}$, which is unlike the model for the error in \citetalias{schlegel1998}, where they assume the linear relation $\sigma_\mathrm{SFD}=fE(B-V)$ (their equation 24) finding that their results are consistent with a value of $f=0.16$.
To compare this to our results, the orange shaded areas in the bottom row of Figure \ref{fig:EBV_recovery} show the area where $E(B-V)_\mathrm{rec}-E(B-V)_\mathrm{SFD}>\sigma_\mathrm{SFD}$\footnote{To the authors' knowledge, the only other maps to have errors measured or described are those by \citet{green2015,green2018,green2019}, but the errors from these are rather uninformative for this work, as they are consistent with a value of $0.030$ mag at every resolution, independent of $E(B-V)$.}.
As \nhigh\ is close to the resolution of the \citetalias{schlegel1998} map, this is the more straightforward comparison, which shows that our method provides a smaller scatter at $E(B-V)\gtrsim0.1$ mag, but it is larger for smaller $E(B-V)$ values.

Due to how \citetalias{schlegel1998} combined the DIRBE and ISSA maps the comparison at \nlow\ is trickier, as the uncertainty at that scale is dominated by the contribution from DIRBE, but they do not explicitly explore the contributions from each to their error model.
If, as a whole, the fractional uncertainties are dominated by DIRBE, they should remain the same between \nhigh\ and \nlow, while if they are dominated by ISSA they should scale inversely with the square root of the pixel area, with the expectation that the real noise of the map at that resolution would be in between these two values.
Furthermore, this model only represents small-scale errors, with the possibility of other systematics such as mismodeling of the zodiacal light and temperature correction becoming dominant by \nlow\ (D. Finkbeiner, private communication).
To compare to our results at \nlow\ we have taken the simple, optimistic assumption that the \citetalias{schlegel1998} errors will scale as the inverse of the square root of the pixel area, which is shown with the lighter orange-shaded region on the bottom left panel.
Under this assumption, the uncertainty in our method is moderately larger than that from the \citetalias{schlegel1998} map, though close enough that a more realistic treatment of their uncertainties could make them roughly equal to ours, as for \nhigh.


\section{Reconstruction of $\delta$ maps using the recovered dust extinction map}\label{sec:delta_maps}

In what follows, we introduce a method to estimate pixelized overdensities, that takes into account dust with the additional advantage of using the intrinsic distributions of galaxy luminosities and colors.
We show that this reduces systematics in the estimated overdensities.
This approach would enable the calculation of individual galaxy weights that take into account the density variations induced by dust in measurements of the galaxy correlation function for cosmological parameter estimations.

When inferring overdensities, $\delta$, we measure the fraction of galaxies bright enough to be observed ($f$) as a function of dust extinction.
If we assume this to be uncorrelated with fluctuations in the original galaxy counts, this can be written as

\begin{align*}
    1+\delta_{\mathrm{LSS},i} &= n_i/\bar{n}\\
    1+\delta_{\mathrm{ext},i} &= n_{\mathrm{ext},i} / \bar{n}_{\mathrm{ext}} = f_i n_i / \bar{n}_{\mathrm{ext}}\\
    \bar{n}_{\mathrm{ext}} &= \langle f_i n_i \rangle = \langle f_i \rangle \bar{n}\\
    \Rightarrow 1+\delta_{\mathrm{ext},i} &= (1+\delta_{\mathrm{LSS},i}) f_i / \langle f_i \rangle 
\end{align*}

\noindent where $n_i$ is the number of galaxies in the voxel $i$ in the absence of dust, $\bar{n}$ is the average dust-free number counts across voxels, $f_i$ the ratio between the post-dust and original galaxy counts in voxel $i$, and the $\mathrm{ext}$ subscript indicates dust-extinguished quantities.
To simplify the use of this relation to correct for MW dust we work with logarithmic values instead of linear ones, as the logarithm decouples $f_i$:

\begin{equation*}
    \log (1+\delta_{\mathrm{ext},i}) = \log (1 + \delta_{\mathrm{LSS},i}) + \log(f_i) - \log(\langle f_i \rangle)
\end{equation*}

\noindent The presence of measurement errors, and possible unaccounted systematics, will induce an error in the recovery process, so in reality, the overdensity that we recover, $\delta_\mathrm{rec}$, will not be an exact match to $\delta_\mathrm{LSS}$.

\begin{figure*}
    \epsscale{1.1}
    \plotone{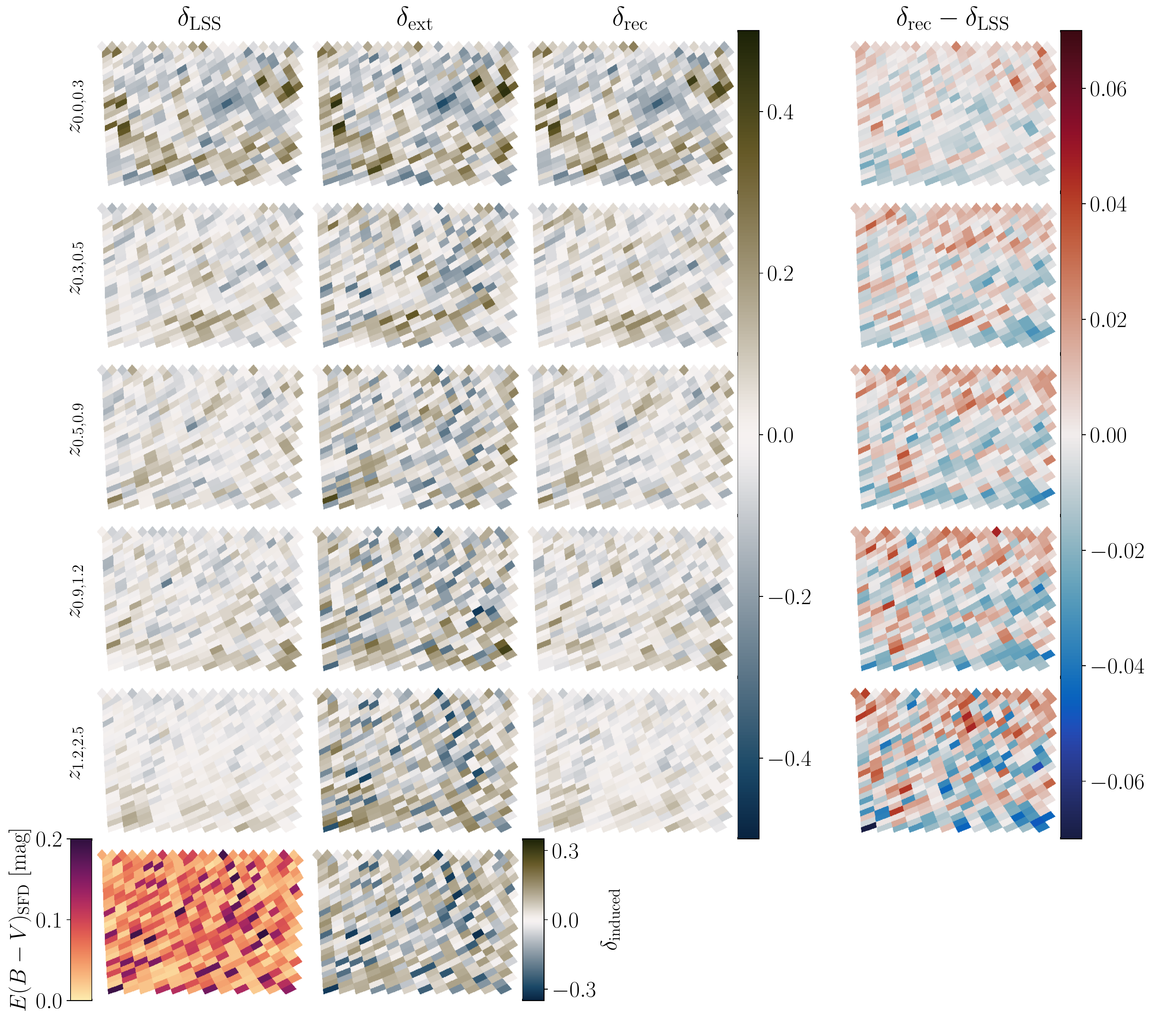}
    \caption{Cartesian view of the input, measured, recovered and residual maps for the overdensity; $\delta_\mathrm{LSS}$, $\delta_\mathrm{ext}$, $\delta_\mathrm{rec}$ and $\delta_\mathrm{rec}-\delta_\mathrm{LSS}$, respectively. The subscript in $\delta_\mathrm{rec}$ denotes that the recovery was performed with the recovered dust map from \S\ref{sec:EBV_recover}. The top five rows each show a different redshift bin, with the last showing the combined results. In the bottom row, the left panel is the same as the bottom middle from Figure \ref{fig:EBV_map_n6}, and the right panel shows the induced signal by that dust screen on a flat ($\delta=0$) overdensity map ($\delta_\mathrm{induced}$).}
\label{fig:delta_map_n6}
\end{figure*}

MW dust creates spurious overdensities that are uncorrelated with the true LSS, as shown in Figure \ref{fig:dust_to_quantities}.
This will lead to increased scatter when measuring LSS without applying any correction.
For this purpose, similar to the method described in \S\ref{sec:EBV_maps}, we propose the following steps to correct LSS measurements for the effect of MW dust:

\begin{enumerate}
    \item Divide the range of redshift covered by the survey into redshift bins, as in \S\ref{sec:EBV_maps}.
    \item Calculate the fraction of remaining galaxies ($f_i$) for a range of values of $E(B-V)$ for each redshift bin.\label{list:method_steps_21}
    \item Divide the area of the sky covered by the survey into angular pixels to generate voxels, as in \S\ref{sec:EBV_maps}.
    \item Calculate the observed overdensity ($\delta_{\mathrm{ext},i}$) using the galaxy counts in each voxel ($n_{\mathrm{ext}}$) and the average galaxy counts across all voxels in the same redshift bin ($\bar{n}_{\mathrm{ext}}$).
    \item For each voxel, calculate $\delta_{\mathrm{LSS},i}$ from $\delta_{\mathrm{ext},i}$, correcting by the $f_i$ value implied from the relation defined in step \ref{list:method_steps_21} and the $E(B-V)$ value of the corresponding pixel of the dust map recovered in \S\ref{sec:EBV_maps}.
    \item Reset the mean value of $\delta$ to 0 by the additive correction of $\log(\langle f_i \rangle)$.
\end{enumerate}

\begin{deluxetable*}{lcccc}
    \tablecaption{Ratio between the standard deviation of $\delta_\mathrm{rec}-\delta_\mathrm{LSS}$ and $\delta_\mathrm{LSS}$, $\sigma_{\delta_\mathrm{rec}-\delta_\mathrm{LSS}}/\sigma_{\delta_\mathrm{LSS}}$, at each redshift and for each dust map used for the recovery.\label{tab:delta}}
    \tablehead{\colhead{}&\multicolumn{2}{c}{\nlow}&\multicolumn{2}{c}{\nhigh}}
    \startdata
             & $E(B-V)_\mathrm{SFD}$ & $E(B-V)_\mathrm{rec}$ & $E(B-V)_\mathrm{SFD}$ & $E(B-V)_\mathrm{rec}$\\
    \hline
    \zlower  & 0.05263 & 0.06538 & 0.11320 & 0.13415\\
    \zlow    & 0.10091 & 0.14430 & 0.14986 & 0.23713\\
    \zmid    & 0.09416 & 0.16873 & 0.16668 & 0.26729\\
    \zhigh   & 0.10667 & 0.20316 & 0.22811 & 0.33348\\
    \zhigher & 0.15733 & 0.40290 & 0.34005 & 0.50962\\
    \enddata
\end{deluxetable*}

We follow a similar logic to the method detailed in \S\ref{sec:dust_vector} to measure the fraction of remaining galaxies (step \ref{list:method_steps_21}).
The bottom panel of Figure \ref{fig:dust_to_quantities} shows the $f$-$E(B-V)$ relations we measure for the synthetic LC.
The differences between redshift bins are on a similar scale to those for $u$ and $u-z$, where the same value of $E(B-V)$ produces changes that vary by a factor of $\sim1.5$ across redshift, with the highest redshift bin being the most affected because of the redshift dependence of the luminosity threshold to enter the sample.
When this is below the characteristic luminosity ($L^\star$) at a given redshift, the number of detected galaxies will change modestly.
If the threshold is above $L^\star$ the number counts are dominated by the faintest detected galaxies, and any change of the luminosity limit significantly affects the number of detections.

For illustrative purposes, in the bottom row of Figure \ref{fig:delta_map_n6}, we show the effect of MW dust on $\delta$.
We chose one of the $E(B-V)$-$\delta$ relations from Figure \ref{fig:dust_to_quantities} (\zhigher), the corresponding input MW dust screen from Figure \ref{fig:EBV_map_n6} (bottom left panel, for convenience repeated in Figure \ref{fig:delta_map_n6}), and we assume an equal number of galaxies ($\delta_{\mathrm{LSS},i}=0$) in all voxels.
We use the $f$-$E(B-V)$ relation to convert the dust map into the false signal ($\delta_\mathrm{induced}$) induced by it.
This is shown on the bottom row, which shows that low dust pixels (light pixels, left panel) induce overdensities (light pixels, right panel), while high dust pixels (dark pixels, left panel) induce under-densities (dark pixels, right panel).

The rest of Figure \ref{fig:delta_map_n6} shows the LSS maps from the simulation ($\delta_\mathrm{LSS}$, left column), the measured maps affected by dust extinction ($\delta_\mathrm{ext}$, middle-left column), the recovered maps ($\delta_\mathrm{rec}$, middle-right column) and the residuals ($\delta_\mathrm{rec}-\delta_\mathrm{LSS}$, right column) from our method at \nlow.
Here we use our recovered $E(B-V)$ map ($E(B-V)_\mathrm{rec}$), as this replicates the error propagation from dust recovery to $\delta$ recovery that would affect our method when applied to real data.
From this, it is clear that, as expected from the results in the bottom row, dust becomes the dominant component of uncorrected $\delta$ measurements at high redshifts, strongly distorting the true distribution, as shown by comparing $\delta_\mathrm{ext}$ at \zhigher\ (middle row, 4$^\mathrm{th}$ from top to bottom) to $\delta_\mathrm{induced}$ (middle row, bottom panel).
From the residuals, it becomes clear that the problem of recovering $\delta$ becomes increasingly harder as a function of redshift.

\begin{figure*}
    \epsscale{1.1}
    \plotone{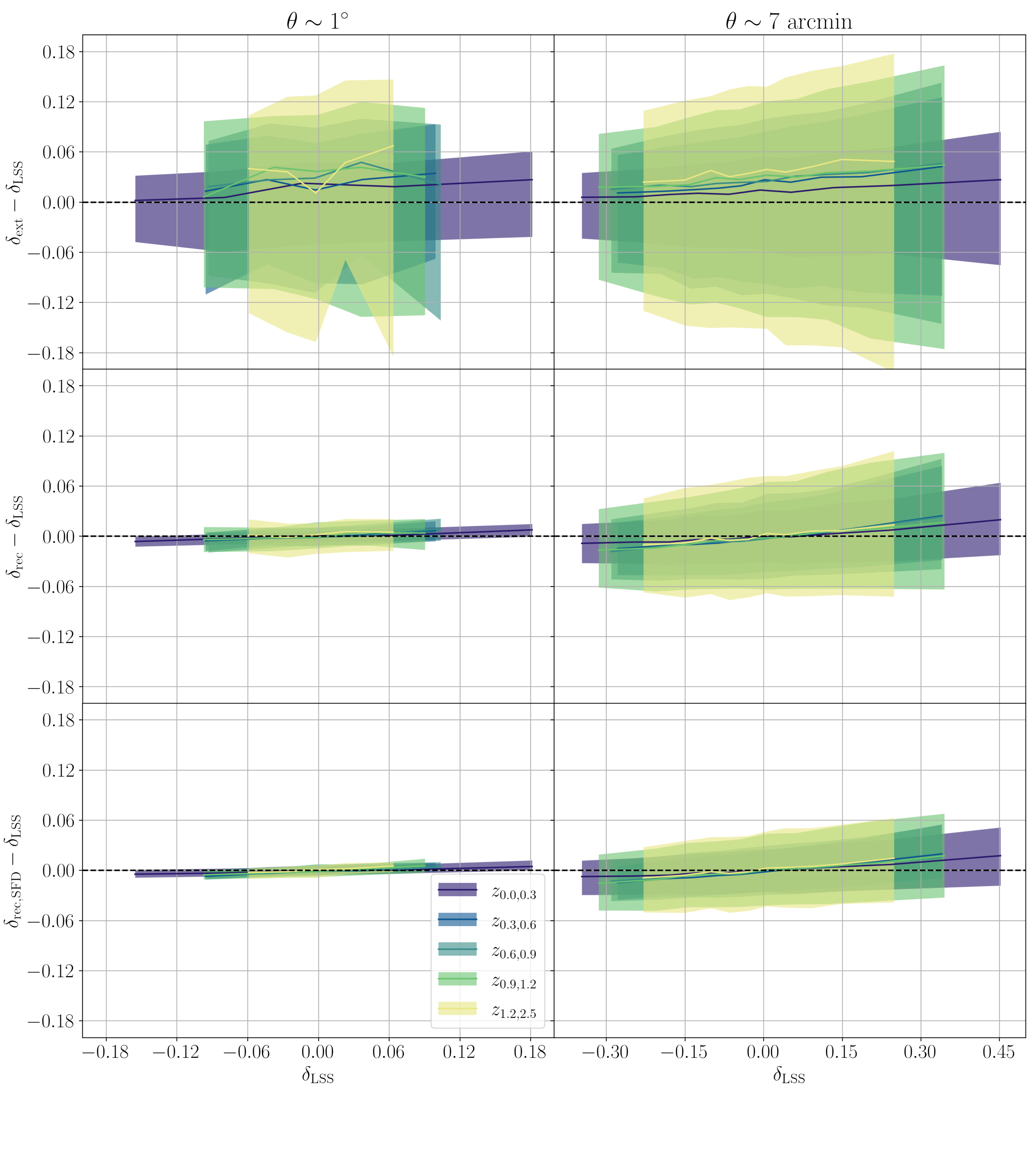}
    \caption{Residuals between recovered values, $\delta_\mathrm{rec}$, after applying the MW dust extinction correction, and input values, $\delta_\mathrm{LSS}$, colored by redshift bin. The solid lines show the running median on bins with equal numbers of galaxies, with shaded regions showing the 16-84\% range. The top row shows the scatter between the dust-extinguished measured overdensities ($\delta_\mathrm{ext}$) and the true LSS ($\delta_\mathrm{LSS}$). The middle row shows the results for our analysis method, which corrects for dust using the recovered map, $E(B-V)_\mathrm{rec}$. The bottom row shows results for an idealized case that re-uses the input map, $E(B-V)_\mathrm{SFD}$, during the recovery. 
    The left and right columns show results for resolutions of \nlow\ and \nhigh, respectively.}
\label{fig:delta_recovery}
\end{figure*}

Figure \ref{fig:delta_recovery} presents the residuals from our method as a function of $\delta_\mathrm{LSS}$ at both \nlow\ and \nhigh.
To serve as a reference, the top row quantifies the effect of dust on the LSS measurements.
The middle row shows the results using the recovered dust map, $E(B-V)_\mathrm{rec}$, with our method achieving a remarkable recovery of $\delta_\mathrm{LSS}$ at \nlow\, with similar scatter across all redshift bins, while at \nhigh\ the improvement from our method is not as strong, and the dependence on extinction induced by MW dust on the residuals remains.

Since we know the input and recovered dust maps, we can study how the uncertainties in our MW dust map recovery propagate into our $\delta$ recovery.
The bottom row from Figure \ref{fig:delta_recovery} shows the results for an idealized case where we estimate overdensities using the same dust map ($E(B-V)_\mathrm{SFD}$) that we applied as a screen to our synthetic LC.
Comparison with the top row shows that, not surprisingly, the residuals increase
significantly when using $E(B-V)_\mathrm{rec}$ to recover $\delta$, though the median lines are largely unaffected.
This illustrates that the slope and $y$-intercept offsets present on $E(B-V)$ only translate weakly into biases in $\delta_\mathrm{rec}$.

Table \ref{tab:delta} contains a summary of these results, showing the standard deviations in the residuals ($\delta_\mathrm{rec}-\delta_\mathrm{LSS}$), divided by the standard deviation of the true LSS ($\delta_\mathrm{LSS}$) to quantify the distortion of the LSS signal that remains after our dust correction.
To put these results into context for cosmology, the baryonic acoustic oscillation scale \citep[$\sim100\mathrm{h}^{-1}$Mpc;][]{eisenstein2005} remains well resolved at our lower resolution of \nlow\ up to $z\sim1$, where the BAO scale is $\sim2.5^\circ$.
Moving higher in redshift, the BAO scale drops to $\sim1.3^\circ$ at $z\sim2.5$.
This suggests that the $\delta$ maps produced by our recovery method are viable for BAO measurements.


\section{Discussion}\label{sec:discussion}

Our results indicate the feasibility of measuring MW dust extinction using extragalactic sources in the LSST, and of reconstructing the underlying LSS, though there is still space for improvements.
While the increase in area from pixels of \nhigh\ to \nlow\ is a factor of $\approx60$, which would result in a factor $\approx8$ improvement for white noise, both our dust and overdensity recovery errors decrease by roughly a factor of 4.
This implies that sources of correlated noise (LSS misidentified as dust, and overall bias in our method) are contributing roughly $20\%$ as much as white noise at \nlow\ but become dominant at \nhigh.
This points to the fact that, while the aim of using different redshift bins is to remove any correlation with the LSS from our $E(B-V)$ map, some contamination persists.
Even with the improvements that we propose below, it is reasonable to expect some level of residual correlation, which can make such a map a non-ideal choice for specific uses like systematics deprojection.
In particular, while we do not expect the LSS reconstruction to harm the underlying statistics, we plan to test in future work the effects on cosmological parameter estimation using our method to reconstruct the LSS with the LSST DESC DC2SSS.

While in \S\ref{sec:filters} we referred to Figure \ref{fig:col_col_z_ap} to argue for the use of one color for the MW dust map recovery in this work, there is second-order information to be captured by the inclusion of additional colors, which could further reduce the biases and uncertainties resulting from our method.
We will explore this in more detail in future work, but an obvious question is which combination of filters optimizes the recovery.
For one color the answer is fairly straightforward: a pair of filters that balances wavelength separation with sensitivity.

Another possible modification would be to use the extreme quantiles of the color distributions (e.g., $1\%$ bluest and $1\%$ reddest) instead of/in addition to their median.
The argument for this is that colors run into physical limits at both the blue and red ends, at one end due to becoming insensitive to star formation rate changes (the reason why the red sequence is narrow), and on the other, the bluest galaxies bump up against physical limits on star formation, hence both could prove sensitive to foreground MW dust.

Our mock catalogs assume a perfectly uniform survey.
However, as shown by \citet{awan2016}, survey non-uniformity can have an important effect on LSS measurements, driven by the overlaps in the tessellation of the survey observing strategy on which our method depends, resulting in inhomogeneities in the depth on angular scales similar to our chosen resolutions.
\citet{awan2016} show that a well-chosen dithering strategy can greatly increase survey uniformity, and a first-order correction can be performed to avoid non-uniformity propagating into spurious dust measurements via fluctuations in $\delta$.
However, in the presence of non-uniform survey depth, using a fixed magnitude for galaxy selection will produce an uneven distribution of errors in the observed magnitudes, and using a fixed detection threshold can propagate that unevenness into $\delta$.
Hence it would be worth investigating correlations between survey non-uniformity and the median galaxy brightness and color at the pixel level so that corrections to these properties could be optimized for MW dust measurements.

\begin{figure}
    \epsscale{1.1}
    \plotone{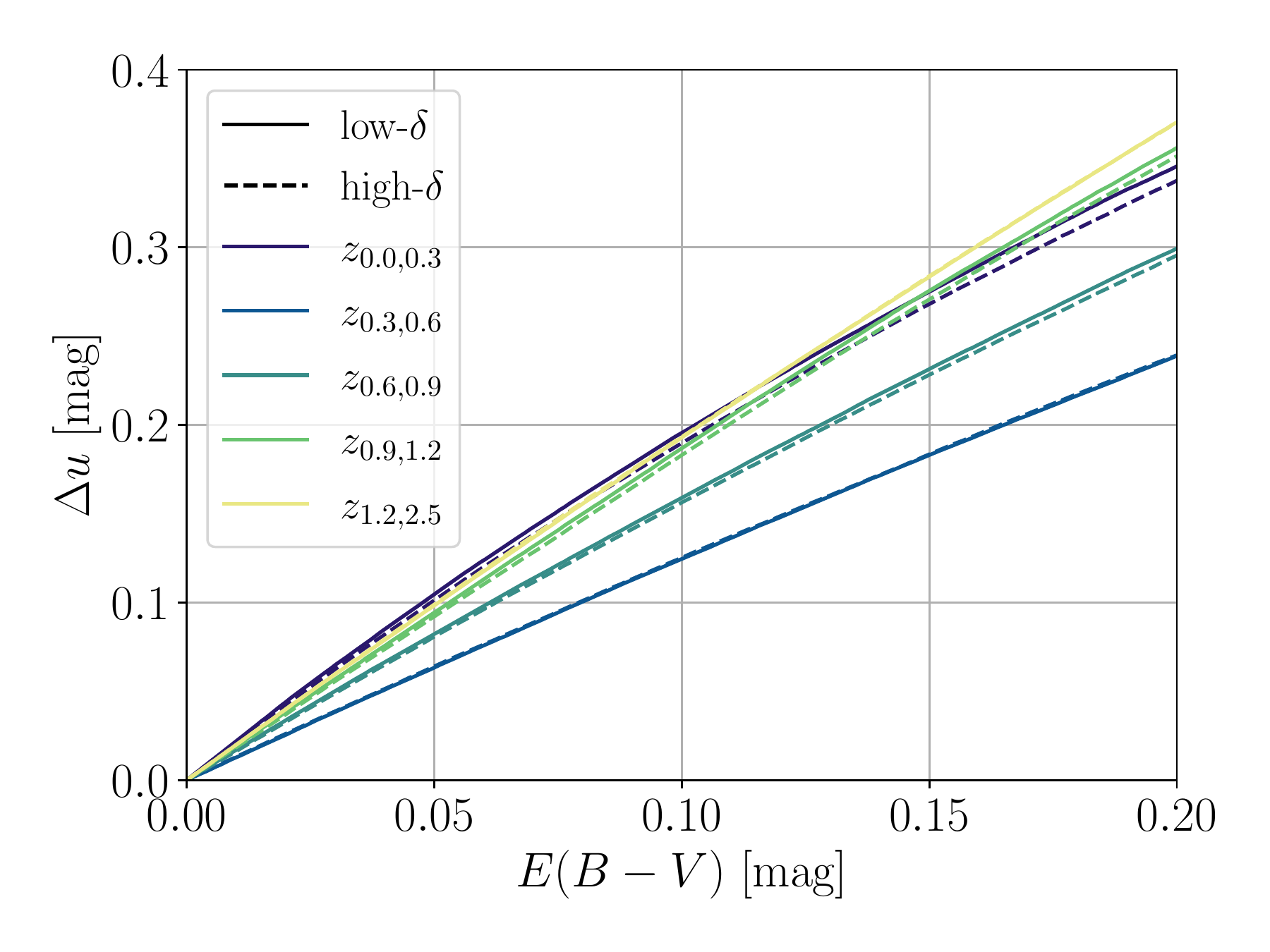}
    \caption{Change in the measured median $u$ magnitude as a function of the added dust screen. Colors as in Figure \ref{fig:dust_to_quantities}. Solid lines show the relation for low-density environments, dashed lines for high-density environments.}
    \label{fig:dust_to_mag_by_delta}
\end{figure}

As galaxy properties correlate with LSS, an obvious suggestion would be to utilize the $E(B-V)$-$f$ relation shown in Figure \ref{fig:dust_to_quantities} to include the initial observations of $\delta$ in our dust recovery method.
That was the original intent for this work, but we found that with our current implementation this leads to a slight increase in the scatter of the recovered quantities, with the residual dust maps showing a clear correlation with the \zhigher\ LSS map.
We emphasize that this result does not rule out including $\delta$ in determining $E(B-V)$ in a future implementation, but in our current method it does not lead to an improvement.

Another possible approach to incorporate $\delta$ would be to measure and apply the $E(B-V)$ vs. $u$ and $E(B-V)$ vs. $u-r$ relations for different environments.
To test this, we updated the calculation used to create Figure \ref{fig:dust_to_quantities}, but this time separating each redshift bin into two sky regions of equal area, with one containing the voxels (at \nlow) above the median number count and the other the rest.
We found no meaningful difference for color or overdensity.
However, for magnitude versus $E(B-V)$, slight differences become visible, as shown in Fig \ref{fig:dust_to_mag_by_delta}.
The small amplitude of these changes leads us to believe that this distinction between low- and high-density environments would be a second-order improvement to our method.

\subsection{Stellar contamination}

As the LSST will detect both stars and galaxies with samples numbering in the billions for each, it is possible to obtain two complimentary dust maps, one from an extension of the work by \citet{schlafly2014,green2015,green2018,green2019} using stars to map the MW dust map, and the other with our method using galaxies.
These two maps would not be fully independent due to star-galaxy contamination, even if there is only a little contamination of stars/galaxies at high/low galactic latitudes.
But since the fluctuations of stars do not follow statistics of $\Lambda$CDM it should be possible to combine the contribution from stars, assumed randomly distributed, and from galaxies to pixels.
The complementarity comes from the fact that our galaxy-based method will provide the best results at high galactic latitudes, where the number of detections is less affected by dust, whereas for a star-based method, due to the sparsity of stars at high latitudes, the best measurements are achieved near the galactic plane.

It should be noted that a classification method that provides good separation between stars and galaxies is critical for either method to work, as \citet{chiang2019} show that stellar-based dust maps are affected by pollution from unresolved extragalactic sources.
By construction, our method should prove immune to this type of bias, but it may well suffer the inverse problem, being polluted by stars.
While this issue requires further testing to be assessed, we expect it to mostly affect the $\delta$ measurement as galaxies and stars share a similar range of intrinsic $u-z$ color, and any bright outlier in $u$ would be a prime candidate for being a misidentified, nearby star.
Any correlation between the amplitude of LSS fluctuations and galactic latitude should reveal residual effects of dust extinction and contamination by stars, and this could be used to estimate whether the separation between stars and galaxies from the classification is sufficient.
The distribution of contamination across our redshift bins will depend on the specific photometric redshift algorithm used.

Another aspect of stellar contamination is the impact of foreground stars upon the completeness of the chosen galaxy sample.
While we do not expect this contamination to dominate galaxy number counts, as the choice to limit the sky area to that where $E(B-V)\lesssim0.2$ means that we are working in a relatively low stellar density regime, our simulations do not include stars to test this effect.
This is another area where using the LSST DESC DC2 simulations, which contain realistic stars and galaxies, would be of benefit to explore this systematic effect.

\subsection{Bayesian approach}

Figure \ref{fig:EBV_recovery} illustrates that our method produces a dust extinction map with uncertainties comparable to those of the emission-based \citetalias{schlegel1998} map and that these maps are complementary because our uncertainties are nearly constant in arithmetic units of $E(B-V)$ whereas the reported \citetalias{schlegel1998} uncertainties are fractional.
This makes \citetalias{schlegel1998} more precise in very low-dust regions and our map more precise in intermediate-dust regions; our advantage may not extrapolate to $E(B-V)>0.2$, as the stellar density is generally higher in such dusty regions, and we have not accounted for stellar contamination of galaxy samples or the reduction in detectable galaxy counts in crowded fields.
This complementarity motivates the use of an existing dust map like \citetalias{schlegel1998} as a prior for our method, which would then simultaneously refine that input dust map while measuring LSS.

Inspired by previous Bayesian methods to estimate the LSS \citep[e.g.,][]{jasche2013,porqueres2019} that marginalize over systematic effects (such as dust), we now propose a method that extends this idea by also adding a posterior for dust.
Our method makes it possible to calculate the likelihood that excursions in galaxy properties ($\vec{D}$, where the vector runs over multiple properties and all redshift bins) at a given sky location are due to a particular combination of Galactic dust extinction, parameterized by $A_V$, and large-scale structure, parameterized by $\vec{\delta}$ (where the vector runs over redshift bins).
This naturally encourages a Bayesian approach where this likelihood is multiplied by prior information from external dust maps to produce a posterior probability distribution for $\vec{\delta}$ and $A_V$ in each pixel as follows:

\begin{equation}
P(\vec{\delta},A_V | \vec{D}) \propto L(\vec{D} \; | \vec{\delta}, A_V) p(\vec{\delta}) p(A_V).
\label{eq:bayesian}
\end{equation}

Uncertainties in the external dust map are accounted for in the width of the prior $p(A_V)$.
Hierarchical Bayesian analysis can be utilized if dust systematics are modeled as hyper-parameters that modify the $A_V$ prior.
The large-scale structure prior, $p(\vec{\delta})$, is not normally seen in this form but follows from the prior probability distribution assumed for cosmological parameters and the survey window function.
At low angular resolutions, it will be roughly lognormal and uninformative, but a careful treatment would be required to incorporate galaxy bias, non-linearity, and non-gaussianity.
Both $\vec{\delta}$ and $A_V$ priors have spatial correlations; in the case of $\vec{\delta}$ these are described by the correlation function within each redshift slice; for $A_V$, this can be either modeled \citep[e.g.,][]{sale2014,green2019} or inferred \citep[e.g.,][]{leike2019}.
These considerations recommend the application of techniques developed to separate Cosmic Microwave Background foregrounds \citep[e.g.,][]{planck2016x}, for which spatial templates are sometimes available, from the intrinsic cosmological signal whose phase information is unknown.

In this work, we have limited ourselves to work with only the filter set used for the LSST, but the inclusion of photometry outside the optical range is also worth exploring.
It is known that overdensities of high-redshift galaxies cause excess far-infrared (FIR) emission versus the average FIR background.
An overdensity of galaxies should cause an increase in both optical and FIR emission, whereas MW dust would increase FIR emission while reducing the number of detected galaxies, which means that the addition of 
an FIR map into our method could add further discriminating power.

As mentioned in \S\ref{sec:EBV_maps}, the details of the chosen dust law are not relevant, as by construction we are using the correct law to recover the dust screen we simulate.
The presence of dust law variations across the sky has been known for several decades \citep{savage1979,cardelli1988}, but how these variations depend on wavelength is still an open matter.
\citet{schlafly2016} show that different dust laws recover different degrees of scattering as a function of wavelength, and their results do not agree with all models.
This presents an interesting and significant challenge for our method, which as presented in this work depends on a good understanding of how to go back and forth between $E(B-V)$ and the extinction values for the adopted bands.

The simplest approach would be to make an informed choice of dust law, with the parametrization that best describes the area of the sky covered by LSST.
A still simple extension would be to produce several maps with a selection of different dust laws and parametrizations for the user of said maps to choose from.
A more ambitious prospect would be to expand our method to also produce its own dust law and parametrization measurements.
This seems conceptually possible, as long as dust extinction remains distinguishable from the intrinsic dispersion of galaxies in a suitable choice of magnitude and color space, but it would  require in-depth testing to assess its viability.
This would also require careful modeling of other sources of spectral contamination, including the zodiacal light.

It will be possible to test most of what we have discussed here by implementing this Bayesian formalism using the LSST DESC DC2SSS, once it has been updated with physically motivated positions for the sub-resolution galaxies.


\section{Summary}\label{sec:summary}

This work investigated the use of statistics from extragalactic surveys as a tool to produce a map of MW dust via its effect on the measured galaxy magnitudes, colors, and overdensities, and the use of this map to determine the underlying large-scale structure.
To assess this, we used a synthetic galaxy lightcone that models the expected observations for LSST, to which we added a simulated MW dust screen.
While the simulations used in this work are not publicly available, all of the code used to implement our method and conduct the analysis shown in this work is available on 
GitHub\footnote{\url{https://github.com/MBravoS/MW_dust_galcat}}.

First, we defined a method to calculate how each measured property varies with $E(B-V)$.
We divided the lightcone into voxels, using both redshift bins and angular pixels at two resolutions ($\sim7'$ and $\sim1^\circ$).
Next, we calculated the overdensity,  median magnitude, and median color of the galaxies in each voxel.
Then we measured the expected uncertainties and compared them to existing dust maps.
Our proposed method appears promising, with a measured standard deviation of $\sim0.005\ \mathrm{mag}$ ($\sim0.015\ \mathrm{mag}$) from our LSST-like LC at a resolution of \nlow (\nhigh), roughly comparable to that from \citet{schlegel1998} ($\sim0.007$ at $\sim7'$).

We then explored a method to use the dust map recovered directly from the galaxy survey to simultaneously recover the true LSS, which would allow taking full cosmological advantage of the unprecedented combination of area and depth of LSST.
We find that this method leads to remarkably small uncertainties at all redshifts, especially at \nlow, with the most striking results being the strong removal of the dust-induced signal that would otherwise dominate the measured LSS at high redshifts.
Our achieved uncertainties on the map of overdensities are a factor of 1.5-2 larger than would be achieved by utilizing the (unknowable) true dust map for the correction, and they are several times smaller than those obtained without an explicit dust correction.

Finally, we discussed aspects that were not considered in this exploratory analysis and proposed potential improvements.
One possible extension is a Bayesian approach that uses an existing dust map as a prior that gets refined via information from the galaxy survey while enabling a simultaneous recovery of the underlying LSS.


\section*{Acknowledgments}
This paper has undergone internal review in the LSST Dark Energy Science Collaboration.
The internal reviewers were David Alonso, Eli Rykoff and John Parejko.
MB led the implementation of the method and analysis of the results.
EG provided the original idea for the method implemented, and contributed to the development of the method and to the text of the paper.
NP contributed to the development of the method and to the text of the paper.
JD and RW developed the simulation used in this work and edited the text of the paper.
We acknowledge the help provided by Pedro Fluxa in the implementation and optimization of the algorithm, and the thorough feedback provided by Hunma Awan on the draft during the Collaboration-Wide Review process.
The DESC acknowledges ongoing support from the Institut National de Physique Nucl\'eaire et de Physique des Particules in France; the Science \& Technology Facilities Council in the United Kingdom; and the Department of Energy, the National Science Foundation, and the LSST Corporation in the United States.
DESC uses resources of the IN2P3 Computing Center (CC-IN2P3--Lyon/Villeurbanne - France) funded by the Centre National de la Recherche Scientifique; the National Energy Research Scientific Computing Center, a DOE Office of Science User Facility supported by the Office of Science of the U.S. Department of Energy under Contract No. DE-AC02-05CH11231; STFC DiRAC HPC Facilities, funded by UK BIS National E-infrastructure capital grants; and the UK particle physics grid, supported by the GridPP Collaboration.
This work was performed in part under DOE Contract DE-AC02-76SF00515.
MB and NP acknowledge the support from BASAL AFB-170002 CATA, CONICYT Anillo-1417 and Fondecyt Regular 1191813.
EG was supported by the Department of Energy grants DE-SC0011636 and DE-SC0010008.
The Geryon/Geryon2 cluster housed at the Centro de Astro-Ingenier\'ia UC was used for the calculations performed in this paper.
The BASAL PFB-06 CATA, Anillo ACT-86, FONDEQUIP AIC-57, and QUIMAL 130008 provided funding for several improvements to the Geryon/Geryon2 cluster.
We acknowledge the use of the Legacy Archive for Microwave Background Data Analysis (LAMBDA), part of the High Energy Astrophysics Science Archive Center (HEASARC).
HEASARC/LAMBDA is a service of the Astrophysics Science Division at the NASA Goddard Space Flight Center.
The Millennium Simulation databases used in this paper and the web application providing online access to them were constructed as part of the activities of the German Astrophysical Virtual Observatory.

\software{astropy \citep{astropy}, cmocean \citep{cmocean}, HEALPix \citep{healpix}, matplotlib \citep{matplotlib}, numpy \citep{numpy}, pandas \citep{pandas}, scipy \citep{scipy}}


\bibliographystyle{aasjournal}
\bibliography{papers}


\appendix

\restartappendixnumbering

\section{Results with a different simulation}\label{app:other_sims}

To compare our synthetic LC to other models we have chosen an existing LC made for the Dark Energy Spectroscopic Instrument \citep[DESI;][]{desi2016}, using the \citet{gonzales2014} version of GALFORM \citep{cole2000}, run on the Millennium MS-W7 simulation \citep{guo2013}.
While this LC includes many additional galaxy properties relevant for DESI that will not be probed directly by LSST (e.g., emission lines), it provides all the properties we require for our analysis in a large enough sky area and depth to compare to our \Buz\ LC.
The main driver for this choice is the abundance of literature exploring and using an LC made with this model \citep[e.g.,][]{cole2000,benson2003,baugh2005,bower2006,font2008,lacey2008,lagos2011b,gonzales2014,lacey2016}, though it is highly relevant to this comparison that GALFORM is known to fail to accurately reproduce the color distribution of galaxies \citep[e.g.,][]{font2008,merson2016,bravo2020}.
Table \ref{tab:sim_comp} shows a brief comparison between these two LCs.

\begin{figure}
    \epsscale{1.1}
    \plotone{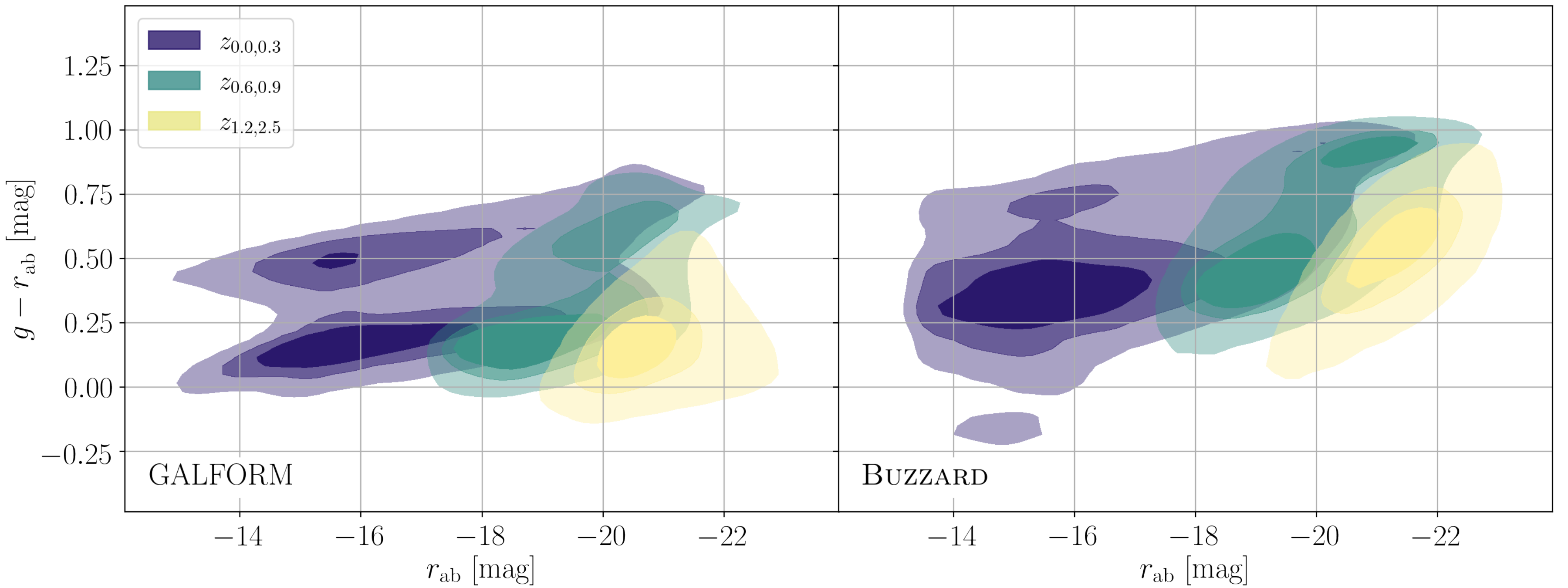}
    \caption{Rest-frame color-magnitude diagrams for GALFORM and \Buz, at the same magnitude limit ($r<24.8$). From darkest to lightest, the contours mark the highest-density region that contains $38.3$, $68.3$, and $95.4\%$ of the galaxies, respectively. Each color corresponds to a different redshift bin.}
\label{fig:mag_col_z_ab}
\end{figure}

We apply the same photometry and redshift errors detailed in \S\ref{sec:synth_cats} (Equations \ref{eq:1} to \ref{eq:3}) to the GALFORM LC, which leads to a small reduction of completeness (from $r\leq25$ to $r\lesssim24.8$).
Since the GALFORM LC does not reach the depth of the \Buz\ highres LC, in this section we have added a sub-sample from \Buz, cut to the same magnitude limit ($r<24.8$).
Figure \ref{fig:mag_col_z_ab} shows a comparison between the rest-frame color-magnitude diagrams for the GALFORM and  \Buz\ highres LCs for three redshift bins (\zlower, \zmid, and \zhigher).

When the same constraints are applied to both LCs, they display a similar trend of splitting into two distinct populations by color, one red and one blue, at low redshift.
There is good agreement in the $r$ magnitude distribution, but it is clear that the two models predict noticeably different colors.
This difference is present across all bins and clearest at low redshift.
The tighter color distribution in GALFORM implies that our dust and overdensity recovery method should perform better using mock catalogs derived from this simulation.

\begin{deluxetable}{lcc}
    \tablecaption{Comparison between the GALFORM-DESI and \Buz-LSST LCs.
    \label{tab:sim_comp}}
    \tablehead{\colhead{}&\colhead{DESI LC}&\colhead{LSST LC}}
    \startdata
    Model name & GALFORM & \Buz\\
    Model type & SAM & SHAM\\
    \# galaxies & $\sim5\times10^6$ & $\sim3\times10^8$\\
    Sky area & $\sim50\ \deg^2$ & $\sim400\ \deg^2$\\
    Redshift limit & $z\leq2.5$ & $z\lesssim8$\\
    Magnitude limit & $r\leq25$ & $r\lesssim29$\\
    \enddata
\end{deluxetable}

The results for the $E(B-V)$ and $\delta$ recoveries are shown in Table \ref{tab:sim_results_comp}.
Except for $E(B-V)$ in the lowest redshift bin (\zlower), GALFORM predicts significantly smaller uncertainties (roughly a factor of 2) for our method compared to \Buz.
Due to the previously mentioned issues with galaxy colors produced by GALFORM at low redshift, we do not expect smaller uncertainties than those shown in Tables \ref{tab:EBV} and \ref{tab:delta}.
However, GALFORM might  turn out to provide a better model 
for galaxy colors at $z>1.5$, where \Buz\ produces inaccurate colors by construction.

\begin{deluxetable}{lcccc}
    \tablecaption{Standard deviation of $\Delta E(B-V)_i=E(B-V)_{\mathrm{rec},i}-E(B-V)_\mathrm{SFD}$ (in magnitudes) and $\Delta\delta_i=\delta_{\mathrm{rec},i}-\delta_{\mathrm{LSS},i}$ of the voxels, for each redshift and the final combined value of $E(B-V)$, for GALFORM and \Buz\ LCs at the same magnitude limit ($r<24.8$).
        \label{tab:sim_results_comp}}
    \tablehead{\colhead{}&\colhead{$\Delta E(B-V)_\mathrm{GALFORM}$}&\colhead{$\Delta E(B-V)_\Buz$}&\colhead{$\Delta \delta_\mathrm{GALFORM}$}&\colhead{$\Delta \delta_\Buz$}}
    \startdata
    \nlow             &         &         &         &        \\
    \zlower           & 0.01907 & 0.01620 & 0.00426 & 0.00817\\
    \zlow             & 0.00786 & 0.00986 & 0.00548 & 0.01115\\
    \zmid             & 0.00744 & 0.01094 & 0.00714 & 0.01558\\
    \zhigh            & 0.00429 & 0.01038 & 0.00975 & 0.02261\\
    \zhigher          & 0.00371 & 0.01579 & 0.01509 & 0.03252\\
    \textbf{combined} & 0.00367 & 0.00627 &         &        \\
    \hline
    \nhigh            &         &         &         &        \\
    \zlower           & 0.07954 & 0.04590 & 0.02721 & 0.03520\\
    \zlow             & 0.05675 & 0.03210 & 0.03075 & 0.04406\\
    \zmid             & 0.03701 & 0.04417 & 0.04184 & 0.06611\\
    \zhigh            & 0.02190 & 0.06455 & 0.05335 & 0.10184\\
    \zhigher          & 0.01613 & 0.14377 & 0.07204 & 0.15369\\
    \textbf{combined} & 0.01447 & 0.02207 &         &        \\
    \enddata
\end{deluxetable}


\end{document}